# Accommodating the Analysis Model in Multiple Imputation for the Weibull Mixture Cure Model: Performance under Penalized Likelihood

Changchang Xu[1,2],
Laurent Briollais[1,2]
Irene L Andrulis[2,3],
Shelley B Bull[1,2] (corresponding author), shelley.bull@utoronto.ca

1. Division of Biostatistics, Dalla Lana School of Public Health, University of Toronto
2. Lunenfeld-Tanenbaum Research Institute, Sinai Health, Toronto, ON, Canada
3. Department of Molecular Genetics, Temerty Faculty of Medicine, University of Toronto

## Abstract

**Introduction** In analysis of time-to-event outcomes, a mixture cure (MC) model is preferred over a standard survival model when the sample includes individuals who will never experience the event of interest. Motivated by a cohort study of breast cancer patients with incomplete biomarkers, we develop multiple imputation (MI) methods assuming a *Weibull* proportional hazards (PH-MC) analysis model with multiple prognostic factors. However, for MI with fully conditional specification, an incorrectly-specified imputation model can impair accuracy of point and interval estimates.

**Objectives and Methods** Our goal is to propose imputation models that are compatible with the Weibull PH-MC analysis models. We derive an exact conditional distribution (ECD) imputation model which involves the analysis model likelihood. Using simulation studies, we compare effect estimate bias and confidence interval (CI) coverage under alternative imputation models including the ECD model, an approximation that includes a cure indicator (cECD), and a comprehensive simple (CS) model. For robust parameter estimation in finite and/or sparse samples, we incorporate the Firth-type penalized likelihood (FT-PL) and combined likelihood profile (CLIP) methods into the MI.

**Results** Compared to complete case analysis, MI with penalization reduces estimation bias and improves coverage. Although ECD and cECD perform similarly at higher event rates, ECD generates smaller bias and higher coverage at lower rates. CS has larger bias and lower coverage than ECD and cECD, but CIs are narrower than for cECD.

**Conclusions** In analyses of biomarkers and composite subtypes for prognosis studies such as in breast cancer, use of compatible imputation models and penalization methods are recommended for MC modelling in samples with low event numbers and/or with covariate imbalance.

## 1 Introduction

In analyzing time-to-event outcomes, such as disease recurrence in studies of prognosis in which subjects have a possibility of being event free, it is uncertain whether a censored patient is cured. The main assumptions of classical survival models are violated due to the existence of those individuals who will not experience the event even when the follow-up time is substantially long. In this circumstance, it is natural to consider mixture cure (MC) models[1, 2].

Letting $T$ be the time to the event of interest, the MC model is defined as

$$\Pr(T > t|\boldsymbol{X}) = p(\boldsymbol{X}) + \left(1 - p(\boldsymbol{X})\right)\ S_0(t|\boldsymbol{X}) \qquad (1).$$

where $p(\mathbf{X}) = \frac{exp(\boldsymbol{\alpha}^\top \mathbf{X})}{1+exp(\boldsymbol{\alpha}^\top \mathbf{X})}$ is the probability of being cured conditional on a vector of covariates $\mathbf{X}$ (i.e., prognostic factors) for incidence, and $S_0(t|\boldsymbol{X}) = exp(-exp(\boldsymbol{\beta}^\top \mathbf{X})t^\gamma)$ is a parametric *Weibull* survival model ($\gamma$ as the shape parameter) for the latent survival distribution at time $t$ among the non-cured, conditional on $\mathbf{X}$. The parameter vectors $\boldsymbol{\alpha}$ and $\boldsymbol{\beta}$ are interpreted as log odds ratios and log hazard ratios respectively. Accordingly, we specify a parametric baseline hazard $h_{00}(t|\boldsymbol{X}) = \gamma t^{\gamma-1}$and hazard function $h_0(t|\boldsymbol{X}) = \gamma exp(\boldsymbol{\beta}^\top \mathbf{X})t^{\gamma-1}$; other baseline functions such as piecewise constant can be considered[3].

The MC framework is appealing because it separately models the probability of being cured and the hazard of an event in the non-cured group; this allows for differentiation of associations of the same factor with outcomes from the two parts of the model (i.e., incidence and latency).

### 1.1 MC analysis for breast cancer prognosis

The methodology we develop is motivated by a prospectively ascertained cohort of axillary lymph-node-negative (ANN) breast cancer patients (*n*=887) followed for up to 15 years[3-7], with greater than 85% remaining recurrence-free. We apply a *Weibull* proportional hazards mixture cure (*Weibull* PH-MC) analysis to investigate the prognostic effect of multiple molecular prognostic biomarkers quantified through Tissue Microarray Analysis (TMA). In practice, the binary biomarker values are used to classify patients into established clinically-relevant subtype categories. In some cases, the TMA procedure yields missing measurements for at least one of the biomarkers leading to an undefined tumour subtype among 34.7% of patients (Figure 1A).

Although Missing Not At Random (MNAR) cannot be excluded, we observed that the occurrence of missing values for the biomarkers depends on other covariates (Section S1.1.3) which is consistent with missing at random (MAR). We adopt multiple imputation (MI) to handle the missing values for covariates in MC analysis, using multiple imputation with chained equations (MICE)[8-12].

Because tumour subtypes are composite, i.e. derived variables defined by biomarkers as constituents, imputing subtypes directly (active imputation) or obtaining subtypes by first imputing biomarkers (passive imputation) can lead to different results.[13,14] In the ANN study, the constituent biomarkers are measured one at a time, and each biomarker has relatively low missing value rates in comparison to the derived categorical subtype (the composite variable) in which missing values accumulate (Figure 1). Furthermore, passive imputation allows for more flexibility when additional biomarkers are incorporated or when subtypes need to be redefined. Clements *et al*[13] suggest passive imputation better preserves the functional relationship between the imputed composite covariate and its constituents. However, imputing missing values of biomarkers instead of derived subtypes may introduce some incompatibility between the analysis and the imputation models, motivating our investigation of an exact conditional distribution (ECD) imputation model[8], i.e., an imputation model that is derived from the joint product of likelihood of the analysis model and conditional probability of the incomplete biomarker given other covariates of the analysis model. In finite samples such as the ANN cohort, the MC model likelihood can be non-quadratic, with ML estimates that are inflated or infinite. Our prior work using the Weibull PH-MC model demonstrated that Firth-type penalized likelihood (FT-PL) methods yield finite and bias-reduced parameter estimates, with smaller likelihood ratio test (LRT) p-values [15,16]. Likewise, a Wald-type confidence interval (CI) usually fails to quantify the precision of the parameters particularly in the presence of low event rates. Applying MI for missing values exacerbates the limitations of ML and Wald-type CI, which we addressed using FT-PL likelihood-based inference.

## 1.2 Survival analysis with missing covariates

### 1.2.1 Multiple imputation model specification

Multiple imputation (MI) is a three-step procedure: (1) statistically impute missing values in the observed dataset via joint modelling (JM) or fully conditional specification (FCS), and repeat to obtain $m$ completed datasets, (2) fit an analysis model in each the $m$ imputed datasets, and (3) combine the results of model fitting across all datasets (Section S1.1.3). Within the first step, specification of the imputation model is a major challenge for proper implementation of MI, because bias can arise when the imputation and analysis models are incompatible, which is closely related to the idea of uncongeniality[8]. Bartlett *et al*[10] developed a general principled method, namely substantive model compatible fully conditional specification (SMC-FCS), to specify an imputation model derived from a joint model for the covariates and the outcome under which both imputation and analysis models are conditional. If $\boldsymbol{X}^p$ and $\boldsymbol{X}^f$ represent the covariates with and without missing values (such that $\boldsymbol{X} = [\boldsymbol{X}^p, \boldsymbol{X}^f]$), and $Y$ is the outcome, then $f(\boldsymbol{X}^p|\boldsymbol{X}^f, Y)$ is proportional to $f(Y|\boldsymbol{X}^p, \boldsymbol{X}^f) * f(\boldsymbol{X}^p|\boldsymbol{X}^f)$. Because FCS MI requires specification of an imputation model for each of the covariates in $\boldsymbol{X}^p$, these authors propose a modification of FCS for multivariate $\boldsymbol{X}^p$, examining the Cox-PH model and other models in detail. This approach has been applied across a range of regression frameworks, including linear, logistic, Cox-PH survival, Cox-PH mixture cure, and cause-specific competing risk survival models.[10-13]

In particular, Beesley *et al*[12] derived an exact conditional distribution (ECD) imputation model for the Cox PH-MC model using the kernel of the conditional density of the incomplete covariate. The development is based on a partial likelihood function and an Expectation-Maximization (EM) algorithm for parameter optimization. Assuming covariates to be conditionally normal or Bernoulli random variables, the authors derive ECD models and propose an approximate ECD (aECD) model to improve computational efficiency. Cure status as a random variable is also included as a predictor and treated as another missing variable in the imputation. Their empirical studies focus on evaluation of various approximations in models with bivariate normal covariates and one binary covariate, and consider event rates of 30-50% with balanced binary covariates. The results suggest that although ECD consistently produces the smallest mean parameter bias, computation can become burdensome with increasing model complexity; the aECD model is less computationally demanding, and bias is modest compared to ECD in most scenarios.

### 1.2.2 Profile likelihood inference for MC analysis model in MI

Studies in logistic regression and in the *Weibull* PH-MC model[15, 16, 17] have revealed that, in sparse or small sample datasets (i.e., few events and/or imbalanced categorical covariates), parameter estimates are subject to finite-sample bias, and data separation produces ML estimation breakdown. In this case, MI with Rubin's Rule[18] (RR) for combining point estimates and variances from multiple-imputed datasets with symmetric confidence intervals (CI's) is not valid. Based on profile likelihood, a combination of penalized likelihood profiles (CLIP) method[15, 17] has improved properties in comparison to RR. In simulation studies of the *Weibull* PH-MC model, CLIP-CIs have higher coverage with narrower CI width, and always produce finite CI endpoints. To our knowledge, the application of CLIP in comparison of the ECD with alternative imputation methods is a novel approach.

In this report, we develop and evaluate imputation models for the *Weibull* PH-MC model with missing covariate values at the individual patient level. Other than the Cox-PH MC model, we are not aware of an ECD approach for any other MC models, including the *Weibull* MC. Therefore, our aim is to derive ECD imputation for the *Weibull* MC model that uses multiple binary covariates as predictors and ensures compatibility between the analysis and imputation models. We consider various imputation models for defining predictors and evaluate the performance of MI for the *Weibull* mixture-cure models in finite samples relevant to studies of multiple binary biomarkers and disease subtypes. In Section 2, we derive the ECD for the *Weibull* PH-MC and describe the proposed imputation models and methods that are evaluated by simulation studies and applied in the motivating prognostic study in Section 3 and 4 respectively. We incorporate penalized likelihood and profile likelihood-based inferences in the MC analysis model to improve estimation efficiency for survival data with low event rates and unbalanced categorical covariates. In Section 3, we report simulation studies designed to evaluate mean bias and CI coverage of effect estimates produced by different imputation model specifications under empirical settings of varying event rate with missing values in the covariates. In Section 4, we demonstrate application of the proposed imputation model(s) in prognostic analysis of the motivating study cohort that uses tumour biomarkers to categorize patients into prognostic subtypes that have clinical relevance.

## 2 Statistical Methods

### 2.1 Specification of the analysis model

For each individual $i$ ($i = 1,2, \ldots, n$) in a sample of size $n$, we observe the event indicator $y_i$ (1 for event, 0 for censored), time to event $t_i$ (event time or observation time), and observed $\boldsymbol{x}_{*,i}$ for the vector of covariates $\boldsymbol{X}$. In the ANN application, the tumour subtype $\boldsymbol{W}$ is a multinomial variable with $\omega + 1$ categories, specified as $\boldsymbol{W} = [W_1, \ldots, W_\omega]$ with binary indicators $W_1, \ldots, W_\omega$ for each of the categories relative to a reference category. $\boldsymbol{W}$ is defined using the biomarker variables $\boldsymbol{X}_{bio}$ ($\boldsymbol{X}_{bio} = [X_1, \ldots, X_q]$ with $X_1, \ldots, X_q$ each assumed to be Bernoulli). Therefore, we define $\boldsymbol{X^p} = [\boldsymbol{W}, \boldsymbol{X}_{d-\omega}]$ ($\boldsymbol{X}_{d-\omega} = [X_{\omega+1}, \ldots, X_d]$ allows for the possibility of a vector of additional biomarkers not used for defining $\boldsymbol{W}$). As shown in Figure 1, the relationships between the $\boldsymbol{W}$ and $\boldsymbol{X}_{bio}$ are defined as

$W_{Her2} = X_{Her2}$,

$W_{TN} = (1 - X_{Her2})(1 - X_{ER})(1 - X_{PR})$,

$W_{LuminalA} = (1 - X_{Her2})(1 - (1 - X_{ER})(1 - X_{PR}))(1 - X_{Ki67})$, and

$W_{LuminalB} = (1 - X_{Her2})(1 - (1 - X_{ER})(1 - X_{PR}))X_{Ki67}$.

We assume $\boldsymbol{w}_i$ are observed values of $\boldsymbol{W}$ included as covariates in the analysis model with $\boldsymbol{\alpha}_{subtype}$ and $\boldsymbol{\beta}_{subtype}$ as the vectors of coefficients, and $\boldsymbol{x}_{d-\omega,i}$ and $\boldsymbol{x}^f_{*,i}$ are the values of the remaining covariates $\boldsymbol{X}_{d-\omega}$ (with $\boldsymbol{\alpha}_{d-\omega}$ and $\boldsymbol{\beta}_{d-\omega}$) and $\boldsymbol{X}^f$ (with $\boldsymbol{\alpha}_f$ and $\boldsymbol{\beta}_f$). With the cure indicator $C$ ($C = 1$ as cured and $C = 0$ as uncured), the MC model has the incidence part:

$$logit(P(C = 1)) = \alpha_0 + \boldsymbol{\alpha}_{subtype}\boldsymbol{W} + \boldsymbol{\alpha}_{d-\omega}\boldsymbol{X}_{d-\omega} + \boldsymbol{\alpha}_f\boldsymbol{X}^f;$$

and the latency part:

$$h(T|C = 0) = \gamma \, exp\big(\beta_0 + \boldsymbol{\beta}_{subtype}\boldsymbol{W} + \boldsymbol{\beta}_{d-\omega}\boldsymbol{X}_{d-\omega} + \boldsymbol{\beta}_f\boldsymbol{X}^f\big)T^{\gamma-1}.$$

The complete data *Weibull* PH-MC likelihood is

$$L(\boldsymbol{\theta}|\boldsymbol{x}_{*,i}) = \prod_{i=1}^{n} \left(h_0(t_i|\boldsymbol{x}_{*,i})S_0(t_i|\boldsymbol{x}_{*,i})(1-\pi_i)\right)^{y_i} \times \prod_{i=1}^{n} \left(\pi_i + (1-\pi_i)S_0(t_i|\boldsymbol{x}_{*,i})\right)^{1-y_i} \quad (4)$$

where $h_0(t_i|\boldsymbol{x}_{*,i}) = \gamma\, t_i{}^{\gamma-1} exp(\boldsymbol{\beta}^{\top}\boldsymbol{x}_{*,i})$

$$= t_i{}^{\gamma-1} exp(\beta_0 + \boldsymbol{\beta}_{subtype}\boldsymbol{w}_i + \boldsymbol{\beta}_{d-\omega}\boldsymbol{x}_{d-\omega,i} + \boldsymbol{\beta}_f \boldsymbol{x}_{*,i}^{f}),$$

$$S_0(t_i|\boldsymbol{x}_{*,i}) = exp\left(-exp(\boldsymbol{\beta}^{\top}\boldsymbol{x}_{*,i})\, H_{00}(t_i)\right)$$

$$= exp\left(-exp(\beta_0 + \boldsymbol{\beta}_{subtype}\boldsymbol{w}_i + \boldsymbol{\beta}_{d-\omega}\boldsymbol{x}_{d-\omega,i} + \boldsymbol{\beta}_f \boldsymbol{x}_{*,i}^{f})\, H_{00}(t_i)\right),$$

$H_{00}(t_i) = t_i{}^{\gamma}$ is the cumulative baseline hazard function, and

$\pi_i = \frac{\exp\,(\boldsymbol{\alpha}^{\top}\,\boldsymbol{X}_{*,i})}{1+\exp\,(\boldsymbol{\alpha}^{\top}\,\boldsymbol{X}_{*,i})} = \frac{\exp\,(\alpha_0+\boldsymbol{\alpha}_{subtype}\boldsymbol{w}_i+\boldsymbol{\alpha}_{d-\omega}\boldsymbol{x}_{d-\omega,i}+\boldsymbol{\alpha}_f\boldsymbol{x}_{*,i}^{f})}{1+\exp\,(\alpha_0+\boldsymbol{\alpha}_{subtype}\boldsymbol{w}_i+\boldsymbol{\alpha}_{d-\omega}\boldsymbol{x}_{d-\omega,i}+\boldsymbol{\alpha}_f\boldsymbol{x}_{*,i}^{f})}$ is the cure probability.

Replacing $\boldsymbol{W}$ in Equation 4 with functions of $\boldsymbol{X}_{bio}$, where $\boldsymbol{x}_{bio,i}$ are the observed values of $\boldsymbol{X}_{bio}$ with coefficients $\boldsymbol{\alpha}_{bio}$ and $\boldsymbol{\beta}_{bio}$, the $h_0(t_i|\boldsymbol{x}_{*,i})$, $S_0(t_i|\boldsymbol{x}_{*,i})$, and $\pi_i$ terms can be re-expressed (details shown in Appendix 6.1):

$$h_0(t_i|\boldsymbol{x}_{*,i}) = \gamma\, t_i{}^{\gamma-1}\exp\,(\beta_0 + \boldsymbol{\beta}_{bio}\boldsymbol{x}_{bio,i} + \boldsymbol{\beta}_{int}\boldsymbol{x}_{\boldsymbol{W},i} + \boldsymbol{\beta}_{d-\omega}\boldsymbol{x}_{d-\omega,i} + \boldsymbol{\beta}_f \boldsymbol{x}_{*,i}^{f}),$$

$$S_0(t_i|\boldsymbol{x}_{*,i}) = exp\left(-exp(\beta_0 + \boldsymbol{\beta}_{bio}\boldsymbol{x}_{bio,i} + \boldsymbol{\beta}_{int}\boldsymbol{x}_{\boldsymbol{W},i} + \boldsymbol{\beta}_{d-\omega}\boldsymbol{x}_{d-\omega,i} + \boldsymbol{\beta}_f \boldsymbol{x}_{*,i}^{f})\, H_{00}(t_i)\right),$$

and $\pi_i = \frac{\exp\,(\alpha_0+\boldsymbol{\alpha}_{bio}\boldsymbol{x}_{bio,i}+\boldsymbol{\alpha}_{int}\boldsymbol{x}_{\boldsymbol{W},i}+\boldsymbol{\alpha}_{d-\omega}\boldsymbol{x}_{d-\omega,i}+\boldsymbol{\alpha}_f\boldsymbol{x}_{*,i}^{f})}{1+\exp\,(\alpha_0+\boldsymbol{\alpha}_{bio}\boldsymbol{x}_{bio,i}+\boldsymbol{\alpha}_{int}\boldsymbol{x}_{\boldsymbol{W},i}+\boldsymbol{\alpha}_{d-\omega}\boldsymbol{x}_{d-\omega,i}+\boldsymbol{\alpha}_f\boldsymbol{x}_{*,i}^{f})}$

The $\boldsymbol{x}_{\boldsymbol{W},i}$ are the observed values for the biomarker cross-products:

$\boldsymbol{X}_{\boldsymbol{W}} = [X_{PR}X_{Her2},\ X_{ER}X_{Her2},\ X_{ER}X_{PR},\ X_{ER}X_{PR}X_{Her2},\ X_{PR}X_{Ki67},\ X_{Ki67}X_{ER},\ X_{Ki67}X_{ER}X_{PR},\ X_{Ki67}X_{PR}X_{Her2},\ X_{Ki67}X_{ER}X_{Her2},\ X_{Ki67}X_{ER}X_{PR}X_{Her2}]$ with coefficients $\boldsymbol{\alpha}_{int}$ and $\boldsymbol{\beta}_{int}$.

## 2.2 Imputation model specification

### 2.2.1 Derivation of MC exact conditional distribution imputation model

To specify an imputation model for each of the partially observed biomarkers $\boldsymbol{X}_{bio}$, assume that $\boldsymbol{X} = [\boldsymbol{X}_{bio}, \boldsymbol{X}_{d-w}, \boldsymbol{X}_{\boldsymbol{W}}, \boldsymbol{X}^{\boldsymbol{f}}]$, such that $\boldsymbol{X}_{bio}, \boldsymbol{X}_{d-w}, \boldsymbol{X}_{\boldsymbol{W}}$ are the vectors of covariates that are only partly observed, and $\boldsymbol{X}^{\boldsymbol{f}} = (X_{d+1}, \ldots X_r)$ are fully observed. Let $X_j$ be the $j$th covariate of $\boldsymbol{X}_{bio}$, and $\boldsymbol{X}_{-j}$ denote all remaining elements of $\boldsymbol{X}$. The joint distribution $f(X_j, \boldsymbol{X}_{-j})$ can be factored as $f(X_j|\boldsymbol{X}_{-j}) * f(\boldsymbol{X}_{-j};\zeta)$ where the joint distribution of $\boldsymbol{X}_{-j}$ will not need to be specified in practice.

Based on the SMC-FCS approach of Bartlett *et al*[10], and following the reasoning of Beesley *et al*[12], we specify an ECD imputation model that is compatible with a *Weibull* PH-MC substantive analysis model, based on imputation of each incomplete $X_j$ in $\boldsymbol{X}_{bio}$. We utilize the joint product of the complete data likelihood function from the *Weibull* PH-MC model and the distribution $f(X_j|\mathbf{X}_{-j})$ to derive the conditional distribution $f(X_j|\mathbf{X}_{-j},Y,T)$ for each partly observed $X_j$. We assume that censoring does not depend on $X_j$ but may depend on other covariates, so we do not need to specify a model for the censoring mechanism to derive the conditional distribution of $X_j$.

Suppose $X_j \sim Bern(\psi_j)$ where $\psi_j = expit(\boldsymbol{\sigma}^\top \mathbf{X}_{-j} + \sigma_\text{j} + \sigma_0)$. Then it follows from

$f(X_j|\mathbf{X}_{-j},Y,T) \propto f(Y,T \mid X_j,\mathbf{X}_{-j}) \times f(X_j|\mathbf{X}_{-j})$ that the conditional density function is

$$P(X_j = 1 \mid Y,T,\mathbf{X}_{-j}) \propto L(\boldsymbol{\theta}| X_j = 1,\boldsymbol{X}_{-j}) \times f(X_j = 1 \mid \boldsymbol{X}_{-j};\sigma_0,\sigma_\text{j},\boldsymbol{\sigma})$$

For a sample of $n$ individuals $(i = 1,2,\dots,n)$, the MC likelihood, conditional on $X_{j,i} = 1$, is

$$L_i(\boldsymbol{\theta}| X_{j,i} = 1,\boldsymbol{X}_{-j,i}) = \left[h_0(t_i| X_{j,i} = 1,\boldsymbol{X}_{-j,i})\, S_0(t_i| X_{j,i} = 1,\boldsymbol{X}_{-j,i})(1-\pi_{ij})\right]^{y_i} \times \left[\pi_{ij} + (1-\pi_{ij})S_0(t_i| X_{j,i} = 1,\boldsymbol{X}_{-j,i})\right]^{1-y_i} \quad (5)$$

where $\pi_{ij} = \frac{\exp(\alpha_0+\alpha_j+\boldsymbol{\alpha}_{-j}\boldsymbol{X}_{-j,i})}{1+\exp(\alpha_0+\alpha_j+\boldsymbol{\alpha}_{-j}\boldsymbol{X}_{-j,i})}$ and $f(X_{j,i} = 1 \mid \boldsymbol{X}_{-j,i};\sigma_0,\sigma_\text{j},\boldsymbol{\sigma}) = \psi_{ij}$ for $X_{j,i} = 1$.

We also define $\pi_{ij0} = \frac{\exp(\alpha_0+\boldsymbol{\alpha}_{-j}\boldsymbol{X}_{-j,i})}{1+\exp(\alpha_0+\boldsymbol{\alpha}_{-j}\boldsymbol{X}_{-j,i})}$ and $f(X_{j,i} = 0 \mid \boldsymbol{X}_{-j,i};\sigma_0,\boldsymbol{\sigma}) = 1-\psi_{ij}$.

Then $P(X_{j,i} = 1|y_i,t_i,\boldsymbol{X}_{-j,i}) \propto L_i(\boldsymbol{\theta}| X_{j,i} = 1,\boldsymbol{X}_{-j,i}) \times \psi_{ij} = \mathcal{L}_i(\boldsymbol{\theta},\boldsymbol{\sigma},\sigma_\text{j},\sigma_0,\zeta|X_{j,i} = 1)$ and $P(X_{j,i} = 0|y_i,t_i,\boldsymbol{X}_{-j,i}) = 1-\mathcal{L}_i(\boldsymbol{\theta},\boldsymbol{\sigma},\sigma_\text{j},\sigma_0,\zeta|X_{j,i} = 1) \propto 1-L_i(\boldsymbol{\theta}| X_{j,i} = 1,\boldsymbol{X}_{-j,i}) \times \psi_{ij}$

$= \mathcal{L}_i(\boldsymbol{\theta},\boldsymbol{\sigma},\sigma_\text{j},\sigma_0,\zeta|X_{j,i} = 0) \propto L_i(\boldsymbol{\theta}| X_{j,i} = 0,\boldsymbol{X}_{-j,i}) \times (1-\psi_{ij})$.

As detailed in Appendix 6.2, the conditional density function in logit scale

$$logit(P(X_{j,i} = 1|y_i,t_i,\mathbf{X}_{-j,i}) = log\left(\frac{\mathcal{L}_i(\boldsymbol{\theta},\boldsymbol{\sigma},\sigma_\text{j},\sigma_0,\zeta|X_{j,i}=1)}{1-\mathcal{L}_i(\boldsymbol{\theta},\boldsymbol{\sigma},\sigma_\text{j},\sigma_0,\zeta|X_{j,i}=1)}\right) = log\left(\frac{\mathcal{L}_i(\boldsymbol{\theta},\boldsymbol{\sigma},\sigma_\text{j},\sigma_0,\zeta|X_{j,i}=1)}{\mathcal{L}_i(\boldsymbol{\theta},\boldsymbol{\sigma},\sigma_\text{j},\sigma_0,\zeta|X_{j,i}=0)}\right)$$

$$\propto \quad \log\left(\frac{L_i(\boldsymbol{\theta}|\, X_{j,i}=1, \boldsymbol{X}_{-j,i}) \times \psi_{ij}}{L_i(\boldsymbol{\theta}|\, X_{j,i}=0, \boldsymbol{X}_{-j,i}) \times (1-\psi_{ij})}\right) \tag{6}$$

reduces to

$$logit(P(X_{j,i} = 1|y_i, t_i, \mathbf{X}_{-j,i}) = y_i\beta_j - y_i e^{\beta_j} + (y_i + 1) \log\left\{A(\boldsymbol{X}_{-j,i})e^{B(\boldsymbol{X}_{-j,i})t^{\gamma}}\right\} - (y_i + 1) \log\left\{e^{\alpha_j}A(\boldsymbol{X}_{-j,i})e^{B(\boldsymbol{X}_{-j,i})t^{\gamma}}\right\} + \log\{e^{\alpha_j}A(\boldsymbol{X}_{-j,i})\} - \log\{A(\boldsymbol{X}_{-j,i})\} + \boldsymbol{\sigma}^{\top}\boldsymbol{X}_{-j} + \sigma_j + \sigma_0,$$

$$where \;\; A(\boldsymbol{X}_{-j,i}) = \exp(\alpha_0 + \boldsymbol{\alpha}_{-j}\boldsymbol{X}_{-j,i}) \text{ and } B(\boldsymbol{X}_{-j,i}) = \exp(\beta_0 + \boldsymbol{\beta}_{-j}\boldsymbol{X}_{-j,i}),$$

and can be further simplified as

$$y_i\beta_j - y_i e^{\beta_j} + (y_i + 1)\left(\frac{e^{\alpha_j}e^{\sum_{k\neq j}^{r}\bar{X}_k\alpha_k}e^{e^{\beta_j}e^{\sum_{k\neq j}^{r}\bar{X}_k\beta_k}t^{\gamma}}}{e^{\alpha_j}e^{\sum_{k\neq j}^{r}\bar{X}_k\alpha_k}e^{e^{\beta_j}e^{\sum_{k\neq j}^{r}\bar{X}_k\beta_k}t^{\gamma}}+1}\left[1 + \sum_{k\neq j}^{r}(\bar{X}_k - X_{k,i})\alpha_k + e^{\beta_j}e^{\sum_{k\neq j}^{r}(\bar{X}_k - X_{k,i})\beta_k}\right] - \frac{e^{\sum_{k\neq j}^{r}\bar{X}_k\alpha_k}e^{e^{\sum_{k\neq j}^{r}\bar{X}_k\beta_k}t^{\gamma}}}{e^{\sum_{k\neq j}^{r}\bar{X}_k\alpha_k}e^{e^{\sum_{k\neq j}^{r}\bar{X}_k\beta_k}t^{\gamma}}+1}\left[1 + \sum_{k\neq j}^{r}(\bar{X}_k - X_{k,i}) + e^{\sum_{k\neq j}^{r}(\bar{X}_k - X_{k,i})\beta_k}\right]\right) + \frac{e^{\alpha_j}e^{\sum_{k\neq j}^{r}\bar{X}_k\alpha_k}}{1+e^{\alpha_j}e^{\sum_{k\neq j}^{r}\bar{X}_k\alpha_k}}\left[1 + \sum_{k\neq j}^{r}(\bar{X}_k - X_{k,i})\alpha_k\right] - \frac{e^{\sum_{k\neq j}^{r}\bar{X}_k\alpha_k}}{1+e^{\sum_{k\neq j}^{r}\bar{X}_k\alpha_k}}\left[1 + \sum_{k\neq j}^{r}(\bar{X}_k - X_{k,i})\alpha_k\right] + \boldsymbol{\sigma}^{\top}\mathbf{X}_{-j} + \sigma_j + \sigma_0 + constant \quad .$$

By taking a linear combination of the terms in the above conditional density function in the logit scale (Equation 6), the ECD imputation model for the sampling process under Gibbs Sampler (Section 2.2.5) can be algebraically derived as

$$logit\left(f(X_j^e)\right) = logit\left(P(X_j^e = 1)\right) \sim Y + \exp(\boldsymbol{X}_{-j}) * H_{00}(T) + Y * \boldsymbol{X}_{-j} + \boldsymbol{X}_{-j} \tag{7}$$

where $X_j^e$ and $X_j^o$ are the missing and observed observations for covariate $X_j$, and $H_{00}(T) = T^{\gamma}$ is the cumulative baseline hazard from the parametric *Weibull*. When the imputation model contains derived covariates $\boldsymbol{W}$ that depend on $\boldsymbol{X}_{bio}$, this is considered active imputation. Given the relationship between $\boldsymbol{W}$ and $\boldsymbol{X}_{bio}$ in Section 2.1 and further substitution of $\boldsymbol{W}$ by $\boldsymbol{X}_{bio}$, such that $\boldsymbol{X} = [\boldsymbol{X}_{bio}, \boldsymbol{X}_{d-w}, \boldsymbol{X}_{\boldsymbol{W}}, \boldsymbol{X}^{\boldsymbol{f}}]$ and $\boldsymbol{X}_{-j} = [\boldsymbol{X}_{bio_{-j}}, \boldsymbol{X}_{d-w}, \boldsymbol{X}_{\boldsymbol{W}_{-j}}, \boldsymbol{X}^{\boldsymbol{f}}]$, where $\boldsymbol{X}_{\boldsymbol{W}_{-j}}$ does not contain any cross-products with $X_j$, (or $\boldsymbol{X} = [\boldsymbol{X}_{bio}, \boldsymbol{X}_{\boldsymbol{W}}, \boldsymbol{X}^{\boldsymbol{f}}]$ when $\boldsymbol{X}_{d-w} = \emptyset$) and $\boldsymbol{X}_{-j} = [\boldsymbol{X}_{bio_{-j}}, \boldsymbol{X}_{\boldsymbol{W}_{-j}}, \boldsymbol{X}^{\boldsymbol{f}}]$ when subtypes are defined by all biomarkers), Equation 7 for our ANN case can be further specified as

$$logit\left(f\left(X_j^e\right)\right) = logit\left(P\left(X_j^e = 1\right)\right)$$

$$\sim Y + \exp\left(\boldsymbol{X}_{bio_{-j}} + \boldsymbol{X^f} + \boldsymbol{X}_{\boldsymbol{W}_{-j}}\right) * H_{00}(T) + Y * \left(\boldsymbol{X}_{bio_{-j}} + \boldsymbol{X^f} + \boldsymbol{X}_{\boldsymbol{W}_{-j}}\right) + \boldsymbol{X}_{bio_{-j}} + \boldsymbol{X^f} + \boldsymbol{X}_{\boldsymbol{W}_{-j}} \quad (8)$$

Note that if $X_j$ is a biomarker from $\boldsymbol{X}_{d-w}$ with the remaining biomarkers as $\boldsymbol{X}_{d-w_{-j}}$, the ECD model as a counterpart of Equation 8 will be expressed as

$$logit\left(f\left(X_j^e\right)\right) = logit\left(P\left(X_j^e = 1\right)\right)$$

$$\sim Y + \exp\left(\boldsymbol{X}_{bio} + \boldsymbol{X}_{d-w_{-j}} + \boldsymbol{X^f} + \boldsymbol{X_W}\right) * H_{00}(T) + Y * \left(\boldsymbol{X}_{bio} + \boldsymbol{X}_{d-w_{-j}} + \boldsymbol{X^f} + \boldsymbol{X_W}\right) + \boldsymbol{X}_{bio} + \boldsymbol{X}_{d-w_{-j}} + \boldsymbol{X^f} + \boldsymbol{X_W}$$

In Equation 8, the cross-product terms in $\boldsymbol{X}_{\boldsymbol{W}_{-j}}$ can be highly sparse (such that no additional information is provided), and not all cross-products are clinically meaningful (i.e., do not define subtypes). We therefore proceed with a passive imputation model and ignore all the interaction terms between covariates of $\boldsymbol{X}$. The ECD imputation model is then specified by

$$logit\left(f\left(X_j^e\right)\right) = logit\left(P\left(X_j^e = 1\right)\right)$$

$$\sim Y + \exp\left(\boldsymbol{X}_{bio_{-j}} + \boldsymbol{X^f}\right) * H_{00}(T) + Y * (\boldsymbol{X}_{bio_{-j}} + \boldsymbol{X^f}) + \boldsymbol{X}_{bio_{-j}} + \boldsymbol{X^f} \quad (9)$$

Passive imputation has the advantages of better preservation of the functional relationship between the biomarkers and subtypes (as described in Section 1.1) and flexibility to input new biomarkers or consider different subtyping. We implement the imputation via "mice",[20] specifying a logistic regression model "logreg" for each of $\boldsymbol{X}_{bio_{-j}}$, $\boldsymbol{X^f}$ and $Y$, then "polyreg" for each of $Y * (\boldsymbol{X}_{bio_{-j}} + \boldsymbol{X^f})$, and predictive mean matching "pmm" for the elements of $exp\left(\boldsymbol{X}_{bio_{-j}} + \boldsymbol{X^f}\right) * H_{00}(T)$. In each of the imputed datasets in the MICE procedure outlined in Section 2.2.5, the tumour subtype variables are derived from the imputed biomarkers in the same way as in the original dataset (Figure 1). We follow the same procedure for the imputation models in Sections 2.2.2 - 2.2.4).

### 2.2.2 Approximation to exact conditional distribution with cure fraction (cECD)

In a previous study[12], Beesley *et. al.* proposed an imputation model that approximates the ECD model and performs similarly with less computational demand; the approximation is expressed as the linear combinations of all covariates in first order form as in ECD and the interaction of cure probability with all covariates and outcome variables. By analogy with Beesley, we propose a cure-ECD model (cECD) for *Weibull* PH-MC, by including $\boldsymbol{X}_{bio_{-j}}$, $\boldsymbol{X^f}$, $Y$, $(\boldsymbol{X}_{bio_{-j}} + \boldsymbol{X^f})*$ $H_{00}(T)$ and the latent $C$ as predictors,. The imputation model we apply for partly observed $X_j^e$ has the form

$$logit\left(f\left(X_j^e\right)\right) = logit\left(P\left(X_j^e = 1\right)\right)$$

$$\sim C * Y + \mathrm{C} * H_{00}(T) * (\boldsymbol{X}_{bio_{-j}} + \boldsymbol{X^f}) + C * (\boldsymbol{X}_{bio_{-j}} + \boldsymbol{X^f}) + \boldsymbol{X}_{bio_{-j}} + \boldsymbol{X^f} + C \qquad (10)$$

For implementation, we first impute $C$ in a logistic model with predictors $\boldsymbol{X}_{bio_{-j}}$, $\boldsymbol{X^f}$, $Y$ and $T$, ensuring the $C$ value is consistent with $Y$, i.e., for a patient $i$ who had an event ($Y_i = 1$), the imputed cure status is not cured ($C_i = 0$). We subsequently impute $X_j^e$ via "mice" by specifying "logreg" for each of $\boldsymbol{X}_{bio_{-j}}$, $\boldsymbol{X^f}$ and $C$, "polyreg" for $Y * C$ and $C * (\boldsymbol{X}_{bio_{-j}} + \boldsymbol{X^f})$, and "pmm" for each of $C * H_0(T)$ and $C * H_0(T) * (\boldsymbol{X}_{bio_{-j}} + \boldsymbol{X^f})$.

### 2.2.3 Comprehensive Simple (CS) model

A common practical approach used for specifying the imputation model for a MC model is to assume a linear combination of the predictors for imputing missing values in $X_j$ which are $\boldsymbol{X}_{bio_{-j}}$, $\boldsymbol{X}_f$ and the two outcomes (i.e., $T$ and $Y$) in the first order form (not derived from a particular likelihood). Therefore, to impute the missing values of a binary biomarker, $X_j^e$, we apply

$$logit\left(f\left(X_j^e\right)\right) = logit\left(P\left(X_j^e = 1\right)\right) \sim \boldsymbol{X}_{bio_{-j}} + \boldsymbol{X^f} + T + Y \qquad (11)$$

This imputation approach is named *comprehensive simple* (CS), as it includes all the biomarkers used in deriving the tumour subtype covariates, as well as the other covariates (e.g. traditional prognostic factors such as menopausal status and tumour size) and outcome variables from the analysis model, but in the simplest form (i.e., first order form) without interaction terms. The imputation is implemented via “mice” by specifying “logreg” for the binary variables ($\boldsymbol{X}_{bio_{-j}}, \boldsymbol{X}^{\boldsymbol{f}}, Y$) and “pmm” for $T$.

### 2.2.4 Mis-Specified (MIS) model

A less comprehensive approach for specifying the imputation model is to include only the biomarkers ($\boldsymbol{X}_{bio}$) that are used in deriving the tumour subtype covariates of the analysis model as well as $T$ and $Y$, which assumes that the missingness is not associated with any factors other than the biomarkers and the outcome variables. For instance, to impute the missing values of a binary biomarker, $X_j^e$, we apply

$$logit\left(f\left(X_j^e\right)\right) = logit\left(P\left(X_j^e = 1\right)\right) \sim \boldsymbol{X}_{bio_{-j}} + T + Y \qquad (12)$$

where $\boldsymbol{X}_{bio_{-j}}$ includes all the biomarkers except for $X_j$. We name this imputation approach *Mis-Specified* (MIS), because it is a strong assumption to assume missingness in the biomarkers does not depend on any other factors (e.g. tumour size may affect sample availability or IHC assay reagent sensitivity). Unless the missing pattern is MCAR, a mis-specified imputation model such as MIS can produce a more biased estimate than CS. For example, if information is not available for some TPFs associated with missing biomarker values, then the missing data mechanism is MNAR and parameter estimation may be biased and inefficient. The implementation here is also through “mice”, specifying “logreg” for the binary biomarkers ($\boldsymbol{X}_{bio_{-j}} Y$) and “pmm” for $T$.

### 2.2.5 MICE procedure

The MICE procedure has been implemented in several software packages (i.e., R and Stata) that allow specification of the conditional distribution for each partly observed variable and then utilize these distributions to impute variables one-by-one in an iterative process[19,20]. Assume $f\left(\phi_j | \boldsymbol{X}, Y, T\right)$ is the posterior distribution of $\phi_j$. To impute missing values for $(X_1^o, \dots X_d^o)$, we apply the iterative chained equations algorithm in ‘mice’, following the final imputation model in 2.2.1. At iteration $t$, the procedure updates imputed values by a Gibbs sampler that successively draws

$$\phi_1^{*(t)} \sim f\left(\phi_1 | X_1^{o,(t-1)}, X_2^{(t-1)}, \ldots, X_d^{(t-1)}, \boldsymbol{X^f}, Y, T\right)$$
$$X_1^{*e(t)} \sim f\left(X_1 | X_2^{(t-1)}, \ldots, X_d^{(t-1)}, \phi_1^{*(t)}, \boldsymbol{X^f}, Y, T\right)$$
$$\phi_2^{*(t)} \sim f\left(\phi_2 | X_1^{(t)}, X_2^{o,(t-1)}, \ldots, X_d^{(t-1)}, \boldsymbol{X^f}, Y, T\right)$$
$$X_2^{*e(t)} \sim f\left(X_2 | X_1^{(t)}, X_3^{(t-1)}, \ldots, X_d^{(t-1)}, \phi_2^{*(t)}, \boldsymbol{X^f}, Y, T\right)$$
$$\ldots\ldots$$
$$\phi_d^{*(t)} \sim f\left(\phi_d | X_1^{(t)}, X_2^{(t)}, \ldots, X_{d-1}^{o,(t-1)}, \boldsymbol{X^f}, Y, T\right)$$
$$X_d^{*e(t)} \sim f\left(X_d | X_1^{(t)}, X_2^{(t)}, \ldots, X_{(d-1)}^{(t)}, \phi_d^{*(t)}, \boldsymbol{X^f}, Y, T\right).$$

The iteration proceeds until convergence or a pre-specified maximum number is reached. A single imputed dataset is obtained with the above iteration process, and the process is then repeated multiple ($m$) times to produce multiple-imputed datasets (Figure S1). In a standard MICE procedure, the imputation model is specified based on the type of the partly observed variable. For instance, a logistic regression model ("logreg") is specified for a binary covariate $X_j$ that is assumed to follow a Bernoulli distribution, with $\mathbf{X}_{-j}$, $Y$, and $T$ as predictors. For imputing numerical covariates, we use predictive mean matching ("pmm") as implemented in 'mice'.

## 2.3 Profile likelihood inference for MI analysis model

### 2.3.1 Firth-type penalized likelihood for parameter estimation

Given the likelihood function for MC in Equation 4 (and Equation S.1)[16], the common approach for regression parameter estimation is via ML to solve the score equations $\frac{\partial logL(\boldsymbol{\theta})}{\partial\boldsymbol{\theta}} = \frac{\partial l(\boldsymbol{\theta})}{\partial\boldsymbol{\theta}} = U(\boldsymbol{\theta}) = \mathbf{0}$, where $\mathbf{0}$ is a vector of the same length as $\boldsymbol{\theta}$. The modified score equations based on the proposed Firth-type penalization can be expressed as the following form:

$$U^*(\theta) \equiv \begin{bmatrix} U_\alpha \\ U_\beta \\ U_\gamma \end{bmatrix} + \frac{1}{2} \times \begin{bmatrix} V_{tr\alpha} \\ V_{tr\beta} \\ V_{tr\gamma} \end{bmatrix} \quad (13)$$

where $U_{\boldsymbol{\alpha}}$ and $U_{\boldsymbol{\beta}}$ are vectors of score equations ($U(\boldsymbol{\theta})$) for coefficient parameters from the incidence part and the latency part, and $V_{tr\boldsymbol{\alpha}}$ and $V_{tr\boldsymbol{\beta}}$ are vectors of $tr\left(I(\boldsymbol{\theta})^{-1} \times \frac{\partial I(\boldsymbol{\theta})}{\partial \theta_j}\right)$ for coefficient parameters from the two parts of the model; $U_\gamma$ and $V_{tr\gamma}$ are score equation and $tr\left(I(\boldsymbol{\theta})^{-1} \times \frac{\partial I(\boldsymbol{\theta})}{\partial \gamma}\right)$ respectively for parameter $\gamma$. Detailed expressions are shown in Supplemental S1.2. Under the assumption that the maximum penalized likelihood estimates exist, which satisfies $U^*(\boldsymbol{\theta}^*) = 0$, iterative adjustment (i.e., a Newton-type algorithm) can be applied in the penalized score function to obtain optimized (and bias reduced) estimates:

$$\boldsymbol{\theta}^*_{(it+1)} = \boldsymbol{\theta}^*_{(it)} + I^{-1}\big(\boldsymbol{\theta}_{(it)}\big)\boldsymbol{U}^*(\boldsymbol{\theta}^*_{(it)}) \qquad (14)$$

### 2.3.2 CLIP confidence interval and hypothesis test

Suppose the likelihoods ($l(\boldsymbol{\theta})$ and $l^*(\boldsymbol{\theta})$) for each imputed data are expressed as in Equation S1 and Equation S2. Let $l^{(k)}_{\theta_j}$ be the $k$th imputed-data logarithmic (or penalized) likelihood function, and $l^{(k)}_p(\theta_j)$ be the $k$th imputed-data logarithmic profile (penalized) likelihood function corresponding to $\theta_j$ that is maximized at $\left(\hat{\theta}^{(k)}_j, \hat{\theta}^{(k)}_{-j}\right)$. We also denote the $k$th imputed-data logarithmic (penalized) likelihood ratio statistic as $D^{(k)}(\theta_j) = 2\{l^{(k)}\left(\hat{\theta}^{(k)}_j, \hat{\theta}^{(k)}_{-j}\right) - l^{(k)}_p(\theta_j)\}$, with the signed root $r^{(k)}_{\theta_j}(\theta_j) = sign\left(\theta_j - \hat{\theta}^{(k)}_j\right)\sqrt{D^{(k)}(\theta_j)}$. Since $D$ (or $D^*$ for FT-PL) is assumed to have an asymptotic $\chi^2$ distribution with $df = 1$, $r^{(k)}(\theta_j)$ follows a standard normal conditional on the imputed data[15], denoted as

$$F^{(k)}(\theta_j) = \Phi\{r^{(k)}_{\theta_j}(\theta_j)\} \qquad (15)$$

where $\Phi(.)$ is the standard normal CDF. For the pooled estimates across the multiply imputed data, the average of $F^{(k)}(\theta_j)$ across $m$ imputes,

$$\bar{F}(\theta_j) = \frac{1}{m}\sum_{k=1}^{m} F^{(k)}(\theta_j) \qquad (16)$$

is considered as an approximation to the actual posterior, and the pooled $(1-\alpha)\%$ CI can be obtained as continuous set of values, s.t., $\bar{F}(\theta_j) \geq \alpha/2$ and $\bar{F}(\theta_j) \leq 1-\alpha/2$[15]. For hypothesis testing of the pooled parameter estimate, we utilize the same signed root of cCDF, $\bar{F}(\theta_j)$, and obtain the p-values as its tail values approximated by standard normal distribution. Similar to RR, the ultimate pooled estimate is the average of $m$ imputes, $\left(\hat{\alpha}_j^{(k)}, \hat{\beta}_j^{(k)}\right)$.

## 2.4 Predicted recurrence free survival by MC with MI

For predicting the risk of recurrence in ANN patient data with missing covariate values, we estimate the vectors of $\widetilde{\boldsymbol{\alpha}}$, $\widetilde{\boldsymbol{\beta}}$ and $\tilde{\gamma}$ via a MC model as in Equation 1 with MI. As before, the risk of recurrence at time $t_l$ $(l = 1, \ldots)$ for an individual with $\boldsymbol{X} = \boldsymbol{X_i}$ (a vector of specific values of covariates $\boldsymbol{X}$ for an individual $i$) can be obtained from

$$\Pr\left(T > t_l \middle| \boldsymbol{X} = \boldsymbol{X_i}, \widetilde{\boldsymbol{\alpha}}, \widetilde{\boldsymbol{\beta}}, \tilde{\gamma}\right) = p(\boldsymbol{X_i}|\widetilde{\boldsymbol{\alpha}}) + \left(1 - p(\boldsymbol{X_i}|\widetilde{\boldsymbol{\alpha}})\right) S_0(t_l|\boldsymbol{X_i}, \widetilde{\boldsymbol{\beta}}, \tilde{\gamma}) \qquad (17)$$

where $p(\boldsymbol{X_i}|\widetilde{\boldsymbol{\alpha}}) = \frac{exp(\widetilde{\boldsymbol{\alpha}}^\top \boldsymbol{X_i})}{1+exp(\widetilde{\boldsymbol{\alpha}}^\top \boldsymbol{X_i})}$ and $S_0\left(t_l|\boldsymbol{X} = \boldsymbol{X_i}, \widetilde{\boldsymbol{\beta}}, \tilde{\gamma}\right) = exp\left(-exp\left(\widetilde{\boldsymbol{\beta}}^\top \boldsymbol{X_i}\right) t_l^{\tilde{\gamma}}\right)$. The confidence band of the predicted risk curve is acquired by substitution in the above model with the vector of $(\boldsymbol{\alpha}^{(k)}, \boldsymbol{\beta}^{(k)}, \gamma^{(k)})$ from $m$ imputed datasets that produces the 2.5$^{th}$ (lower) and 97.5$^{th}$ (upper) percentile probability of recurrence $\Pr\left(T > t_l \middle| \boldsymbol{X} = \boldsymbol{X_i}, \widetilde{\boldsymbol{\alpha}}^{(\boldsymbol{lo})}, \widetilde{\boldsymbol{\beta}}^{(lo)}, \tilde{\gamma}^{(lo)}\right)$ and $\Pr\left(T > t_l \middle| \boldsymbol{X} = \boldsymbol{X_i}, \widetilde{\boldsymbol{\alpha}}^{(\boldsymbol{up})}, \widetilde{\boldsymbol{\beta}}^{(up)}, \tilde{\gamma}^{(up)}\right)$ at time $t_l$.

# 3 Simulation studies

## 3.1 Simulation methods

### 3.1.1 Full data generation

***Generating covariates and survival outcomes*** We first generate four IHC biomarkers (i.e., HER2, ER, PR, and Ki67) and two TPFs (tumour size (TUMCT) and menopausal status (MENS0)) as binary variables for each patient according to the observed distributions in the original ANN data: 8%, 71%, 56% and 60% positivity for HER2, ER, PR and Ki67, as well as 41% pre/peri-menopausal status (vs. post), and 45% tumour size greater than 2cm (vs. less than 2cm). According to binary expression (positive or negative) of four IHC biomarkers, we define four intrinsic subtypes[4]: Her2, Luminal A, Luminal B, and Triple Negative (TN). Table S2 reports the realization of biomarker distributions and subtype distributions. We generate the survival outcomes (i.e., probability of cure, and time to recurrence) under two MC models with different specifications of covariates: i) tumour subtypes with MENS0 and TUMCT, and ii) tumour subtypes only. Cure status ($C$, such that $C$=1 for cured and $C$=0 for not cured), RFS time ($T$), and censoring indicator ($Y$, such that $Y = 1$ for observed event/recurrence and $Y = 0$ for censored) are generated for patients based on a MC model. The specific methods to generate survival outcomes and correlated multiple binary covariates follow those of Xu et al[16].

***Models and notation*** The target event rate of 25% is set by adjusting the intercept values $\alpha_0$ and $\beta_0$ to produce an event number around 75 after dropping all cases with missing values for a complete case (CC) analysis. We set the coefficient values for the log odds ratios ($\alpha_j$) and log hazard ratios ($\beta_j$) of the tumour subtypes to be close to the estimates after fitting the same MC model to the ANN TMA data. Binary indicators for subtypes Her2, Luminal A and TN are defined by biomarkers as in Figure 1 and section 2.1. The higher prevalence Luminal B subtype with a relatively high event rate is assumed as the reference category.

The simulation scenarios are generated under two models. Model 1 is a 5-covariate MC model (including the three binary tumour subtype indicator variables and the two binary TPFs), with the incidence and latency parts specified as:

$$logit\big(P(C = 1|W_{Her2}, W_{LuminalA}, W_{TN}, X_{MENS0}, X_{TUMCT})\big) = \alpha_0 + \alpha_{Her2}W_{Her2} + \alpha_{LuminalA}W_{LuminalA} + \alpha_{TN}W_{TN} + \alpha_{MENS0}X_{MENS0} + \alpha_{TUMCT}X_{TUMCT};$$

and:

$$h(T|W_{Her2}, W_{LuminalA}, W_{TN}, X_{MENS0}, X_{TUMCT}, C = 0) = \gamma exp(\beta_0 + \beta_{Her2}W_{Her2} + \beta_{LuminalA}W_{LuminalA} + \beta_{TN}W_{TN} + \beta_{MENS0}X_{MENS0} + \beta_{TUMCT}X_{TUMCT})T^{\gamma-1}.$$

"Model 2" is a 3-covariate tumour subtype only model. Simulations under both models are generated with sample size *n*=1000. The parameter values specified for the two models are summarized in Table 1. Subsequently, the cure variable $C$ and outcome variables ($T$,$Y$) are generated according to the coefficient values.

### 3.1.2 Imposing missingness

To obtain datasets with missing values in the covariates, we first impose missing values in the four TMA biomarkers and subsequently derive tumour subtypes. The 'simsem' R package is used to generate MCAR patterns and MAR patterns, with the missing percentages for the four biomarkers as in the ANN data: i.e., average percentages 30%, 15%, 15%, and 24.7% for Her2, ER, PR and Ki67, respectively. MAR is generated by imposing the missing indicators based on the observed values of TUMCT and MENS0. The detailed methods are described in Xu[15].

Under Model 2, MNAR is also generated by imposing the missing indicators with the same relationships as the above, but we exclude the two TPFs from the analysis model (assuming both are unobserved). The realizations of missing percentage across simulation replicates are shown in Table S3. Tables S4 & S6 report distributions of event rates in full data and CC data for Model 1 and Model 2 simulations, respectively.

### 3.1.3 Missing data analyses

***Complete case definition*** The tumour subtype categories (Her2, Luminal A, Luminal B and TN) are defined by combinations of the binary biomarkers. Therefore, complete case (CC) datasets can be obtained by dropping subjects with a particular biomarker missing at each hierarchical level when defining tumour subtype (Figure 1A). This differs from a "list-wise deletion"[12] approach which excludes all subjects with any missing biomarker values before assigning subtypes. In what follows, we compare the missing data methods under the MI analyses with CC and full data analysis.

***Imputation model specification*** Under both simulation models, we impute missing biomarker values for each individual. For Model 1 simulations, we consider four imputation models (i.e., ECD, cECD, CS and MIS as defined in Section 2), and then fit the correctly specified MC analysis model (covariates are tumour subtype indicators and two TPFs) to the imputed data to generate estimates for MI-ECD, MI-cECD, MI-CS and MI-MIS. For Model 2 simulations, we consider two imputation models, a CS model (commonly used but not most efficient) and a MIS model (likely very biased), and then fit the correctly specified MC analysis models (covariates are tumour subtypes only) to the imputed data to generate estimates for MI-CS and MI-MIS.

***Evaluation of results*** The MI missing data approaches are compared to full data and CC analysis, with parameter estimation by FT-PL. Hypothesis testing and interval estimation are via profile likelihood approaches (PLCI for full data and CC, and CLIP-CI for MI analyses). We report FT-PL parameter estimates across simulation replicates with event rate 25% and 10% (Figure 2), and show 2D mean distance (of $\alpha$ and $\beta$ associated with the same covariate) between FT-PLEs and their underlying parameter values (Figure 3). MSE for the FT-PL estimates of the simulation replicates are reported in supplemental (Table 5S). The CI coverage rate and 95% CI width from simulation replicates are reported under 25% and 10% event rates (Figures 4 and 5). The supplemental materials also contain: mean bias comparison by ML and FT-PL of Model 2 based parameter estimates at 25% event rate; and 2-sided CI coverage comparison among MI imputation models as well as full data and CC by ML and FT-PL estimation methods under 25% or 10% event rate.

## 3.2 Parameter estimation comparison

### 3.2.1 Model 1 based MCAR and MAR

For data generated under Model 1 with 25% event rate, ECD and cECD yield smaller mean estimation bias than CS and MIS models either marginally (Figure 2, Table S5) or jointly (Figure 3), indicating that inclusion of $H_0(T)$ and interactions between **X** and $T$ or $Y$ in the ECD and cECD models tend to produce less biased estimates for both $\alpha$ and $\beta$ parameters. This is consistent with previous findings[12], in which ECD and cECD models are compared under a Cox-PH MC regression model in simulations of event rate $> 30\%$. CC yields the largest mean bias under FT-PL for Her2 and TN parameters (Figure 2), and largest mean bias under ML for all parameters (Table S5); this suggests that FT-PL alone reduces some estimation bias produced by sample size shrinkage and event number reduction in CC but also improves estimation with MI. The parameter estimates across simulation replicates tend to have similar variation over different MI models (Table S5), indicating that differences among mean estimation bias results from imputation model bias rather than unstable estimation. Comparing MCAR and MAR, there is no noticeable difference among the relative performance (estimation bias or MSE) of various MI methods (Table S5).

For data generated under Model 1 with 10% event rate, ECD produces smaller mean bias than cECD consistently in all parameters, both marginally (Figure 2) and jointly (Figure 3). Similar to observations under 25% rate, the MSE ratios (compared to full data) of the two methods are fairly close (Table 5S), which again indicates that ECD yields less biased estimates than cECD with similar empirical variance of estimation values across simulation replicates. MI-cECD, specified to approximate ECD, adopts **X** alone in the first order form whereas ECD takes a higher order form of **X** (i.e., $exp(\mathbf{X})$). Previous studies have shown that adding the cure fraction to the imputation model improves estimation efficiency (bias and relative variance)[8] for event rate >30%. In our study, although cECD includes the cure fraction as additional information compared to ECD, imputation results of cECD seem to be more biased under a lower event number or are more sensitive to data sparseness.

#### 3.2.2 Model 2 based MCAR and MNAR

For data generated with MNAR under Model 2, the missingness in the biomarkers depends on the values of menopausal status and tumour size which are assumed to be unobserved variables for the analysis. The mean bias of estimates of MIS (a mis-specified imputation model that contains the subtypes as the only predictors) is larger than that of CS (an imputation model containing both all the subtypes from the analysis model and the unobserved TPFs). However, MIS bias is much smaller than CC and has a smaller MSE (Figure S2, Table S7). In contrast, CC under MCAR is less biased on average (Table 7S). This suggests that a mis-specified imputation model, missing key variables directly responsible for the missing values, can nevertheless provide better results than CC.

### 3.3 CI coverage rate comparison

For data generated with 25% event rate, the PLCI coverage rates of Model 1 parameters in the full data are close to nominal 95%. For some parameters, ECD, cECD and CS tend to produce slightly higher CLIP-CI coverage than the full data PLCI (Figure 4, Table S8). MI with CLIP-CIs tends to generate higher coverage than nominal value, as observed in prior empirical studies under MC analysis[15] and CLIP-CI application in logistic regression[17], which has been attributed to larger CI width. In the current study, high coverage also seems to be related to larger CLIP-CI width compared to full data (Figure 5). At the expense of computation time, coverage closer to 95% may be achieved as the number of imputations increases significantly (i.e., *m*=100, Table S10). Among the four MI methods, CLIP-CI coverage does not differ noticeably, with MI-ECD coverage slightly higher than the other methods. With a smaller sample size, CC tends to generate more biased estimates with larger MSE. The CC PLCI coverage rates are consistently lower than nominal, as are those of most MI methods despite the high average CI widths, due to highly inflated CC CI endpoints. Between MCAR and MAR, no significant difference is observed among the missing data methods.

For data generated with 10% event rate, both ECD and cECD yield CLIP-CI coverage higher than the PLCI coverage of full data (Figure 4, Table S9), and higher than their counterparts in data generated with 25% rate. With similar CI widths, MI-ECD yields overall higher coverage than MI-cECD, likely a result of less bias in MI-ECD estimates, given that the CLIP-CIs are slightly wider (Figure 5). Coverage for MAR is close to that for MCAR for most parameters; higher coverage for $\beta_{LuminalA}$ and $\beta_{TN}$, may be explained by the dependence of Luminal A and TN subtypes on multiple rather than single biomarkers. Compared to FT-PL, ML produces slightly higher over-coverage.

### 3.4 Summary

To conclude our comparisons among various MI models in parallel to full data and CC, with parameter estimation by FT-PL, and CI coverage by CLIP-CI, we offer the following insights. Compared to ML, FT-PL consistently produces less biased mean estimates with smaller MSE especially when the event rate is low. Consistent with the findings of Beesley *et al*[12] who considered an event rate of $> 30\%$, when the sample has a relatively high event rate (25%), inclusion of interaction terms in the imputation models MI-ECD and MI-cECD produces estimates with very small bias. Under MAR or MCAR with individual variable missing rates from 15% to 30%, MI-ECD bias is smaller than MI models with greater mis-specification (i.e. MI-CS and MI-MIS). At a lower event rate of 10%, MI-ECD retains the least biased result and is more resistant to data sparseness than MI-cECD. Compared to CC, MI with specified models that include at least the biomarkers and $T$ and $Y$ tend to yield less biased estimates.

MI with CLIP-CI ($m = 20$) has coverage rates higher than nominal level and higher than PLCI coverage in full data. Among the MI models, MI-ECD and MI-cECD have similarly higher coverage under 25% event rate, followed by MI-CS. When the event rate is 10%, high coverage of MI-ECD does not seem to be compromised by the reduced event rate, perhaps due to its CI width, while MI-cECD exhibits less over-coverage. The coverage of MI models at both event rates reflects the degree of bias in mean estimates, i.e., MI-ECD with smallest mean bias yields the highest coverage rate in data with 10% event rate.

## 4 Application of MI models in ANN data

Recall that our motivation for the development of new MI methods is a cohort of axillary lymph-node-negative (ANN) breast cancer patients prospectively ascertained from eight Toronto hospitals, between September 1987 and October 1996[4]. At the time of diagnosis, information regarding traditional prognostic factors (TPFs) was collected for all patients, including menopausal status, tumour size, tumour histologic grade, and lymphatic invasion. Recurrence-free survival (RFS) time is defined as the time from diagnosis to confirmation of non-breast recurrence (i.e., distant metastasis). Long-term followup data were collected up to a maximum of 15 years. Tumour samples archived at diagnosis were obtained for 887 patients to measure expression of several breast cancer molecular biomarkers via Tissue Microarray analysis (TMA), and among them, 111 had experienced distant disease recurrence.

***Multivariate analysis*** Following previous studies in the ANN cohort[3,5-7,16], we investigate multiple well-established molecular prognostic biomarkers that are quantified through TMA, including HER2, estrogen receptor (ER), progesterone receptor (PR), and Ki67. The classification of different tumour subtypes as combinations of biomarkers aims to better stratify patients with different prognosis. With four TMA biomarkers, we derive four subtypes (i.e., Her2, Luminal A, Luminal B and TN) to categorize the tumour samples and examine their prognostic association with the risk of recurrence accounting for TPFs. Using the TMA data, measurements of at least one biomarker were missing among a proportion of patients. With association analysis, we identify that some missing biomarker values are related to the observed TPFs values (Table S1) for various reasons (i.e., quality and/or nature of tumour sample, assay reagent sensitivity, etc.), which suggests a pattern of missing at random (MAR). Because different patients have different missing biomarkers, as more biomarkers are used as prognostic covariates in MC regressions, the overall complete case sample size can be markedly smaller. Consequently, we see higher prevalence of i) finite sample estimation bias and non-finite effect estimates in CC analysis; and ii) more extreme estimates from imputed data in MI[14].

***MC modelling conventions*** For analyses that assess tumour subtype differences, we define Luminal B tobe the baseline category: as a frequent subtype with the largest number of recurrence events, models with Luminal B as baseline category are expected to provide more stable and interpretable estimates than those with other subtypes (e.g. Luminal A). Out of practicality and convention[3,15,16], in figures and tables summarizing the MC analyses we report logistic regression coefficients and odds ratios for the long-term probability of being uncured (i.e. ever experiencing disease recurrence), which is the complement probability of cure (as originally formulated). For the latency component we report coefficients and hazard ratios for risk of early recurrence in those who would eventually experience recurrence. Thus, a positive valued estimate of the incidence log-odds associated with a subtype indicates a higher long-term odds of being uncured compared to the reference subtype, and a positive estimate of the latency log hazard ratio corresponds to shorter time to recurrence in those who are uncured.

To demonstrate differences among the multiple MI imputation models, we present the following MC analysis with CC and MI analyses in the ANN study data. In Section 4.1, we fit a multivariate MC model with tumour subtypes and two TPFs (i.e., menopausal status, tumour size) using FT-PL and CLIP inference to explore the sensitivity under varying $m$; meantime, parameter estimation with CLIP inference of MI models also compared three MI imputation models via a comprehensive MC model in which tumour subtypes and four TPFs (i.e., menopausal status, tumour size, histologic grade and presence of lymphatic invasion) are the prognostic factors. In Section 4.2, to further examine differences among imputation models, we apply estimated parameter values to predict overall recurrence free survival (RFS) probabilities for hypothetical individuals with specified subtypes and covariates.

## 4.1 Prognostic effects of tumour subtype and TPFs

Based on empirical evidence from finite sample studies[15,16], we apply MI with FT-PL and CLIP inference methods that are expected to reduce mean bias and improve coverage for multivariate MC analysis in the presence of missing covariates. The 5-covariate analysis model that is the same as Model 1 in the empirical study is first applied to examine parameter estimation sensitivity for different $m$, e.g., 20, 50 and 100, via ECD imputation model (Figure 6A, S6, S7 and Table S11). In this case, the number of imputations is not an important factor for obtaining estimates, as increasing $m$ ($\geq$20) does not lead to much difference for parameter estimates and CLIP-CIs (Figure S7). In the 7-covariate analysis model, the three imputation methods (ECD, cECD and CS), each with $m$=20 MI datasets, yield similar estimates and agree on the level of statistical significance for most of the covariates; an exception is the comparison of TN vs. Luminal B in which the cECD CLIP-CI for the logHR does not exclude the null (Figure 6A). We consider the result from MI-ECD to be the most reliable for an analysis model with a larger number of parameters, as suggested by the simulation studies in which it provides the smallest estimation bias, especially for small event numbers. The CLIP-CIs and p-values of MI-ECD (Figure 6A) indicate that, with adjustment for the effects of TPFs, the Luminal A group shows significantly lower odds ratio (OR) of being uncured and lower hazard ratio (HR) for risk of early recurrence, compared to Luminal B patients; the TN group has a significantly lower relative odds of being uncured and higher HR for early recurrence than Luminal B patients; compared to Luminal B, Her2 patients also have significantly higher HR for early recurrence. MI-ECD suggests significant effects of all four TPFs, both for odds of being uncured and for risk of early recurrence.

Analysis using Luminal A as the reference category gives an equivalent model fit but provides additional subtype comparisons. Compared to Luminal A, Her2 patients have significantly higher odds of being uncured and higher hazard ratio for risk of early recurrence, TN patients have significantly lower odds of being uncured but higher HR for early recurrence, and Luminal B patients have significantly higher odds of being uncured, based on MI-ECD with CLIP-CI (Figure S3). The TPFs have the same estimated ORs and HRs, as changing the reference category for the subtypes does not affect estimation of the other covariates. In general, wider CLIP-CIs are observed for these subtype comparisons, because the lower event number/rate for Luminal A affects the stability of parameter estimation for all subtypes.

The addition of two TPFs (histologic grade and lymphatic invasion) seems to greatly improve the imputation and estimation efficiency, given the CLIP-CI's of the 7-covariate model (Figure 6B) are much narrower than those from the 5-covariate model (Figure 6A). The individual CDF curves (Figure S6) that contributed to the calculation of CLIP-CI suggest histologic grade and lymphatic invasion explain variations in recurrence-free survival and are informative for imputation. This reinforces the finding of highly significant prognostic effect of lymphatic invasion on recurrence probability (OR) in a previous study[3]; and aligns with the observation that the values of histologic grade and lymphatic invasion are associated with the missingness of ER, PR or Ki67 (Table S1). We also assess goodness-of-fit by obtaining AIC for the same analysis model fit to each of $m = 200$ imputed datasets in the 5-covariate model. The overall trend in AIC values across multiple-imputed datasets are MI-ECD < MI-cECD < MI-CS (Figure S5), indicating MI-ECD as the imputation model that produces the better fit.

### 4.2 Predicted recurrence free survival probability over time with MI

To predict the RFS probability of an individual patient over time, we use the FT-PL estimates of log ORs and log HRs in the 5-covariate MC model with three tumour subtype indicators and two TPFs (Figure 6B) via three imputation methods, each with $m$=20 MI datasets. As depicted in Figure 7, we consider two hypothetical individuals: a post-menopausal Her2 patient with large tumour size and a post-menopausal Luminal A patient with a small tumour size, and construct 95% confidence bands around their predicted survival probabilities.

As followup time increases, we observe: i) for all MI methods, the predicted RFS probability loses accuracy, i.e. the CB widens; ii) different MI models lose prediction accuracy over time in different degrees, (i.e., cECD loses accuracy fastest with widest CB), and survival probability predictions vary widely across MI models; iii) the predicted RFS variation among MI models is smaller for the lower susceptibility group (Luminal A) compared to the higher susceptibility group (Her2). The 95% confidence bands are too wide to identify significant differences across the different MI models, which suggests a larger sample size is needed to obtain sufficient precision to make meaningful comparisons.

## 5 Discussion

***Novelty of this paper*** i) The ECD and cECD imputation models for *Weibull* PH MC (being quite distinctive from the Cox-PH MC due to their expressions of likelihood) are derived and empirically validated for the first time. ii) The performance of imputation models (e.g., ECD and cECD) is first-ever compared using penalized likelihood and profile likelihood based-inference, i.e., CLIP-CI and CLIP-test, which demonstrated to be more appropriate than maximum likelihood and Wald-type inference in sparse datasets[15,16]. iii) Unlike existing simulation studies of MI in survival or logistic regression analysis[10,12,21], our empirical study setting is closely based on an existing cohort study.

***Practical simulation setting*** The empirical study we report generates variables according to the characteristics of the motivating cohort of ANN patients. The primary variables follow the marginal and joint distributions observed in the ANN cohort study data, and the missing pattern of biomarkers is also created by observed associations with TPF values, both of which provide a sensible setting for investigating the performance of MI models. In recent studies[12,21], authors generated survival data with two continuous covariates under a fairly high event rate (30~40%) that may be a less realistic observation for certain cancers with low recurrence ($\leq 20\%$). These studies were therefore unrevealing of finite sample bias, data separation and test statistic validity resulting from sparse data. Our investigations include scenarios with lower event rates, and consider categorical variables as the main covariates, settings in which a profile likelihood-based approach is needed for appropriate inferences.

***Parameter estimation for imputation procedure*** Under an ECD model (following the SMC-FCS approach), previous studies used an accept-reject algorithm[12] for imputation model estimation under MC or survival analysis models with non-parametric hazard baseline, which draws from the posterior distribution of the incomplete covariate as the joint product of the likelihood function and the conditional density of the covariate. Although the accept-reject method is intended to provide independent sampling points from a proposed distribution over iterations, the algorithm can be inefficient in computation time and its performance may degrade with increasing numbers of covariates[21]. In the imputation stage for the *Weibull*-PH MC analysis, we apply Gibbs sampling via 'mice' package to obtain the imputation model estimates used to predict missing values for each covariate. As a Markov Chain Monte Carlo (MCMC) method, it generates a sequence of samples by iteratively sampling from the full conditional distribution given the current values of all the other variables[22]. In Gibbs sampling, convergence of the estimates is more efficient, especially when the dimension of covariates is high, and it can be easily adapted through the standard package, even for the most time consuming ECD model.

***Other potential imputation approaches*** During the preliminary screening phase for imputation model selection, several other models also drew our attention. One model[8,12] is a modified version of ECD, which imputes the event time and cure status for all censored individuals, such that those who are not cured will have an exact event time (longer than the observed time) and those who are cured are assumed to have an extremely large value. In current literature,[8] it is not evident that this approach improves the performance of ECD, and it may even perform worse with heavily censored data when the majority of event times are not observed. Another approach that incorporates the analysis model is the stacked approach for MICE, which aims to further improve imputation by ECD[21]. In this strategy, multiple imputation datasets are stacked on top of each other to create a large data set, followed by model fitting in a weighted analysis model. Consequently, the parameter estimates and standard errors are also pooled under a weighted approach. This approach relies on estimating standard errors from individual-level imputed covariate data based on parameter estimates so profile likelihood-based inference of the parameter cannot be incorporated directly. It may therefore be more susceptible to the effects of covariate imbalance, finite-sample bias, and data separation.

***Limitations and future work*** One of the potential limitations of the proposed methods is the computational burden in each MI dataset, as well as in aggregation across datasets for pooled estimates: use of FT-PL for parameter estimation in each MI dataset is more intensive than ML, and calculation of CDFs for CLIP-CI is more intensive than RR-CI for multiply imputed datasets. With the power of high-throughput computation, we can parallelize the process over multiple levels to reduce the imputation burden: i) imputation and estimation processes in each of $m$ datasets, and ii) CLIP-CI estimation process for each of multiple coefficients across imputed datasets. According to the aim of the motivating study to investigate the prognostic value of subtypes derived from biomarkers, we adopted passive imputation. A more general setting that can leverage active imputation may be worth exploring under the proposed imputation methods. When the imputation model includes more variables than those present in the analysis model (e.g. auxiliary variables such as additional biomarkers or administrative meta-data), the imputation model performance could be improved as more potentially relevant information is available[4,6]. Although not specified in our design of ECD and cECD, these auxiliary variables not included in the analysis model, can be included in the imputation model as $\boldsymbol{X}^f$ (as in Equation 9 and 10). Further studies could investigate the potential improvement of imputation efficiency by extending the current imputation models to include auxiliary variables as predictors. Another future direction for investigation is the use of model diagnostics in MC with application of MI, which could also facilitate variable selection for MC models. As briefly considered in Section 4.2 to compare the various imputation models, model goodness of fit ought to be assessed via more established model diagnostic methods including but not limited to AIC (Figure S5), or by Cox-Snell or Martingale residuals[23]. How we can incorporate the FT-PL and profile likelihood-based inference into model diagnostic procedures remains to be explored.

## 6. Appendix

### 6.1 Transformation of Equation 4 by substituting $W$ with $X_{bio}$

The relationships between $\boldsymbol{W}$ and $\boldsymbol{X}_{bio}$ are:

$$W_{Her2} = X_{Her2},$$

$$W_{TN} = (1 - X_{Her2})(1 - X_{ER})(1 - X_{PR}) = (1 - X_{PR} - X_{ER} + X_{ER}X_{PR})(1 - X_{Her2})$$
$$= 1 - X_{Her2} - X_{PR} + X_{PR}X_{Her2} - X_{ER} + X_{ER}X_{Her2} + X_{ER}X_{PR} - X_{ER}X_{PR}X_{Her2},$$

$$W_{LuminalA} = (1 - X_{Her2})\big(1 - (1 - X_{ER})(1 - X_{PR})\big)(1 - X_{Ki67})$$
$$= (1 - X_{Her2})(X_{PR} + X_{ER} - X_{ER}X_{PR})(1 - X_{Ki67})$$
$$= (X_{PR} + X_{ER} - X_{ER}X_{PR} - X_{PR}X_{Her2} - X_{ER}X_{Her2} + X_{ER}X_{PR}X_{Her2})(1 - X_{Ki67})$$
$$= (X_{PR} + X_{ER} - X_{ER}X_{PR} - X_{PR}X_{Her2} - X_{ER}X_{Her2} + X_{ER}X_{PR}X_{Her2} - X_{PR}X_{Ki67}$$
$$- X_{Ki67}X_{ER} + X_{Ki67}X_{ER}X_{PR} + X_{Ki67}X_{PR}X_{Her2} + X_{Ki67}X_{ER}X_{Her2} - X_{Ki67}X_{ER}X_{PR}X_{Her2}),$$

$$W_{LuminalB} = (1 - X_{Her2})\big(1 - (1 - X_{ER})(1 - X_{PR})\big)X_{Ki67}$$
$$= X_{PR}X_{Ki67} + X_{Ki67}X_{ER} - X_{Ki67}X_{ER}X_{PR} - X_{Ki67}X_{PR}X_{Her2} - X_{Ki67}X_{ER}X_{Her2}$$
$$+ X_{Ki67}X_{ER}X_{PR}X_{Her2}.$$
$$= X_{Her2} - X_{PR} + X_{PR}X_{Her2} - X_{ER} + X_{ER}X_{Her2} + X_{ER}X_{PR} - X_{ER}X_{PR}X_{Her2}$$
$$- X_{PR}X_{Ki67} - X_{Ki67}X_{ER} + X_{Ki67}X_{ER}X_{PR} + X_{Ki67}X_{PR}X_{Her2} + X_{Ki67}X_{ER}X_{Her2}$$
$$- X_{Ki67}X_{ER}X_{PR}X_{Her2}.$$

Then for Equation 4, the terms are

$$h_0\big(t_i|\boldsymbol{x}_{*,i}\big) = \gamma\, {t_i}^{\gamma-1} exp\big(\boldsymbol{\beta}^\top \boldsymbol{x}_{*,i}\big) = {t_i}^{\gamma-1} exp\big(\beta_0 + \boldsymbol{\beta}_{subtype}\boldsymbol{w}_{*,i} + \boldsymbol{\beta}_{d-\omega}\boldsymbol{x}_{d-\omega,i} + \boldsymbol{\beta}_f \boldsymbol{x}^f_{*,i}\big)$$
$$= \gamma\, {t_i}^{\gamma-1} \exp\,(\beta_0 + \boldsymbol{\beta}_{bio}\boldsymbol{x}_{bio,i} + \boldsymbol{\beta}_{int}\boldsymbol{x}_{\boldsymbol{W},i} + \boldsymbol{\beta}_{d-\omega}\boldsymbol{x}_{d-\omega,i} + \boldsymbol{\beta}_f \boldsymbol{x}^f_{*,i}),$$

$$S_0\big(t_i|\boldsymbol{x}_{*,i}\big) = exp\left(-exp\big(\boldsymbol{\beta}^\top \boldsymbol{x}_{*,i}\big)\, H_{00}(t_i)\right)$$
$$= exp\left(-exp\big(\beta_0 + \boldsymbol{\beta}_{subtype}\boldsymbol{w}_{*,i} + \boldsymbol{\beta}_{d-\omega}\boldsymbol{x}_{d-\omega,i} + \boldsymbol{\beta}_f \boldsymbol{x}^f_{*,i}\big)\, H_{00}(t_i)\right)$$
$$= exp\left(-exp\big(\beta_0 + \boldsymbol{\beta}_{bio}\boldsymbol{x}_{bio,i} + \boldsymbol{\beta}_{int}\boldsymbol{x}_{\boldsymbol{W},i} + \boldsymbol{\beta}_{d-\omega}\boldsymbol{x}_{d-\omega,i} + \boldsymbol{\beta}_f \boldsymbol{x}^f_{*,i}\big)\, H_{00}(t_i)\right),$$

and $\pi_i = \frac{\exp(\boldsymbol{\alpha}^\top \boldsymbol{X}_{*,i})}{1+\exp(\boldsymbol{\alpha}^\top \boldsymbol{X}_{*,i})}$

$$= \frac{\exp\,(\alpha_0+\boldsymbol{\alpha}_{subtype}\boldsymbol{w}_{*,i}+\boldsymbol{\alpha}_{d-\omega}\boldsymbol{x}_{d-\omega,i}+\boldsymbol{\alpha}_f\boldsymbol{x}^f_{*,i})}{1+\exp\,(\alpha_0+\boldsymbol{\alpha}_{subtype}\boldsymbol{w}_{*,i}+\boldsymbol{\alpha}_{d-\omega}\boldsymbol{x}_{d-\omega,i}+\boldsymbol{\alpha}_f\boldsymbol{x}^f_{*,i})} = \frac{\exp\,(\alpha_0+\boldsymbol{\alpha}_{bio}\boldsymbol{x}_{bio,i}+\boldsymbol{\alpha}_{int}\boldsymbol{x}_{W,i}+\boldsymbol{\alpha}_{d-\omega}\boldsymbol{x}_{d-\omega,i}+\boldsymbol{\alpha}_f\boldsymbol{x}^f_{*,i})}{1+\exp\,(\alpha_0+\boldsymbol{\alpha}_{bio}\boldsymbol{x}_{bio,i}+\boldsymbol{\alpha}_{int}\boldsymbol{x}_{W,i}+\boldsymbol{\alpha}_{d-\omega}\boldsymbol{x}_{d-\omega,i}+\boldsymbol{\alpha}_f\boldsymbol{x}^f_{*,i})}$$

where $\boldsymbol{x}_{W,i}$ are the observed values for the biomarker cross-products:

$\boldsymbol{X}_{\boldsymbol{W}} = [X_{PR}X_{Her2},\ X_{ER}X_{Her2},\ X_{ER}X_{PR},\ X_{ER}X_{PR}X_{Her2},\ X_{PR}X_{Ki67},\ X_{Ki67}X_{ER},\ X_{Ki67}X_{ER}X_{PR},\ X_{Ki67}X_{PR}X_{Her2},\ X_{Ki67}X_{ER}X_{Her2},\ X_{Ki67}X_{ER}X_{PR}X_{Her2}]$ with coefficients $\boldsymbol{\alpha}_{int}$ and $\boldsymbol{\beta}_{int}$.

## 6.2 Complete derivation of the proposed ECD imputation model based on biomarkers ($\boldsymbol{X}_{const}$) in Sec 2.2.2

Expanding the derivation process for the ECD model in Section 2.2.2,

with $B(\boldsymbol{X}_{-j,i}) = \exp\,(\beta_0 + \boldsymbol{\beta}_{-j}\boldsymbol{X}_{-j,i})$, and $A(\boldsymbol{X}_{-j,i}) = \exp\,(\alpha_0 + \boldsymbol{\alpha}_{-j}\boldsymbol{X}_{-j,i})$, as follows:

$$logit(P(X_{j,i} = 1|, y_i, t_i, \mathbf{X}_{-j,i}) = log\left(\frac{\mathcal{L}_i(\boldsymbol{\theta},\boldsymbol{\sigma},\sigma_0,\zeta|X_{j,i}=1)}{\mathcal{L}_i(\boldsymbol{\theta},\boldsymbol{\sigma},\sigma_0,\zeta|X_{j,i}=0)}\right)$$

$$\propto \log\left(\frac{\left(h_0(t_i|X_{j,i}=1)S_0(t_i|X_{j,i}=1)(1-\pi_{ij})\right)^{y_i} \times \left(\pi_{ij} + (1-\pi_{ij})S_0(t_i|X_{j,i}=1)\right)^{1-y_i} \times \psi_{ij}}{\left(h_0(t_i|X_{j,i}=0)S_0(t_i|X_{j,i}=0)(1-\pi_{ij0})\right)^{y_i} \left(\pi_{ij0} + (1-\pi_{ij0})S_0(t_i|X_{j.i}=0)\right)^{1-y_i} \times (1-\psi_{ij})}\right)$$

$$= \log\left\{\left[\gamma e^{\beta_j}B(X_{-j,i})t^{\gamma-1}e^{-e^{\beta_j}B(X_{-j,i})t^\gamma}\left(1-\frac{e^{\alpha_j}A(\boldsymbol{X}_{-j,i})}{1+e^{\alpha_j}A(\boldsymbol{X}_{-j,i})}\right)\right]^{y_i} \times \left[\frac{e^{\alpha_j}A(\boldsymbol{X}_{-j,i})}{1+e^{\alpha_j}A(\boldsymbol{X}_{-j,i})} + \left(1-\frac{e^{\alpha_j}A(\boldsymbol{X}_{-j,i})}{1+e^{\alpha_j}A(\boldsymbol{X}_{-j,i})}\right)e^{-e^{\beta_j}B(X_{-j,i})t^\gamma}\right]^{1-y_i}\right\} - \log\left\{\left[\gamma B(X_{-j,i})t^{\gamma-1}e^{-B(X_{-j,i})t^\gamma}\left(1-\frac{A(\boldsymbol{X}_{-j,i})}{1+A(\boldsymbol{X}_{-j,i})}\right)\right]^{y_i} \times \left[\frac{A(\boldsymbol{X}_{-j,i})}{1+A(\boldsymbol{X}_{-j,i})} + \left(1-\frac{A(\boldsymbol{X}_{-j,i})}{1+A(\boldsymbol{X}_{-j,i})}\right)e^{-B(X_{-j,i})t^\gamma}\right]^{1-y_i}\right\} + \log\left\{\frac{\psi_{ij}}{1-\psi_{ij}}\right\}$$

$$= \log\left\{\left[\frac{\gamma e^{\beta_j}B(X_{-j,i})t^{\gamma-1}e^{-e^{\beta_j}B(X_{-j,i})t^\gamma}}{1+e^{\alpha_j}A(\boldsymbol{X}_{-j,i})} \times \frac{1+e^{\alpha_j}A(\boldsymbol{X}_{-j,i})}{e^{\alpha_j}A(\boldsymbol{X}_{-j,i})+e^{-e^{\beta_j}B(X_{-j,i})t^\gamma}}\right]^{y_i} \times \frac{e^{\alpha_j}A(\boldsymbol{X}_{-j,i})+e^{-e^{\beta_j}B(X_{-j,i})t^\gamma}}{1+e^{\alpha_j}A(\boldsymbol{X}_{-j,i})}\right\} - \log\left\{\left[\frac{\gamma B(X_{-j,i})t^{\gamma-1}e^{-B(X_{-j,i})t^\gamma}}{1+A(\boldsymbol{X}_{-j,i})} \times \frac{1+A(\boldsymbol{X}_{-j,i})}{A(\boldsymbol{X}_{-j,i})+e^{-B(X_{-j,i})t^\gamma}}\right]^{y_i} \times \frac{A(\boldsymbol{X}_{-j,i})+e^{-B(X_{-j,i})t^\gamma}}{1+A(\boldsymbol{X}_{-j,i})}\right\} + \log\left\{e^{\boldsymbol{\sigma}^\top\mathbf{X}_{-j}+\sigma_j+\sigma_0}\right\}$$

$$= \log\left\{\left[\frac{\gamma e^{\beta_j}B(X_{-j,i})t^{\gamma-1}e^{-e^{\beta_j}B(X_{-j,i})t^\gamma}}{e^{\alpha_j}A(\boldsymbol{X}_{-j,i})+e^{-e^{\beta_j}B(X_{-j,i})t^\gamma}}\right]^{y_i} \times \frac{e^{\alpha_j}A(\boldsymbol{X}_{-j,i})+e^{-e^{\beta_j}B(X_{-j,i})t^\gamma}}{1+e^{\alpha_j}A(\boldsymbol{X}_{-j,i})}\right\} - \log\left\{\left[\frac{\gamma B(X_{-j,i})t^{\gamma-1}e^{-B(X_{-j,i})t^\gamma}}{A(\boldsymbol{X}_{-j,i})+e^{-B(X_{-j,i})t^\gamma}}\right]^{y_i} \times \frac{A(\boldsymbol{X}_{-j,i})+e^{-B(X_{-j,i})t^\gamma}}{1+A(\boldsymbol{X}_{-j,i})}\right\} + \boldsymbol{\sigma}^\top\mathbf{X}_{-j} + \sigma_j + \sigma_0$$

$$= \log\left\{\left[\frac{e^{\beta_j} e^{-e^{\beta_j}}\left[A(\boldsymbol{X}_{-j,i}) + e^{-B(X_{-j,i})t^{\gamma}}\right]}{e^{\alpha_j} A(\boldsymbol{X}_{-j,i}) + e^{-e^{\beta_j} B(X_{-j,i})t^{\gamma}}}\right]^{y_i}\right\} - \log\left\{\frac{\left[e^{\alpha_j} A(\boldsymbol{X}_{-j,i}) + e^{-e^{\beta_j} B(X_{-j,i})t^{\gamma}}\right]\left[1 + A(\boldsymbol{X}_{-j,i})\right]}{\left[1 + e^{\alpha_j} A(\boldsymbol{X}_{-j,i})\right]\left[A(\boldsymbol{X}_{-j,i}) + e^{-B(X_{-j,i})t^{\gamma}}\right]}\right\} + \boldsymbol{\sigma}^{\top}\mathbf{X}_{-j} + \sigma_j + \sigma_0$$

$$= \log\left\{e^{y_i \beta_j} e^{-y_i e^{\beta_j}}\right\} - \log\left\{\left[\frac{e^{\alpha_j} A(\boldsymbol{X}_{-j,i}) + e^{-e^{\beta_j} B(X_{-j,i})t^{\gamma}}}{A(\boldsymbol{X}_{-j,i}) + e^{-B(X_{-j,i})t^{\gamma}}}\right]^{y_i+1}\right\} + \log\left[\frac{1 + e^{\alpha_j} A(\boldsymbol{X}_{-j,i})}{1 + A(\boldsymbol{X}_{-j,i})}\right] + \boldsymbol{\sigma}^{\top}\mathbf{X}_{-j} + \sigma_j + \sigma_0$$

$$= y_i\beta_j - y_i e^{\beta_j} - (y_i + 1)\left(\log\left\{e^{\alpha_j} A(\boldsymbol{X}_{-j,i}) e^{e^{\beta_j} B(X_{-j,i})t^{\gamma}} + 1\right\} - \log\left\{A(\boldsymbol{X}_{-j,i}) e^{B(X_{-j,i})t^{\gamma}} + 1\right\}\right) + \log\left[1 + e^{\alpha_j} A(\boldsymbol{X}_{-j,i})\right] - \log\left[1 + A(\boldsymbol{X}_{-j,i})\right] + \boldsymbol{\sigma}^{\top}\mathbf{X}_{-j} + \sigma_j + \sigma_0$$

Following the same operations and assumptions as in Beesley et. al.[12], we treat terms that do not depend on $\boldsymbol{X}_i$ as constant and $\log(1+z) \approx \log(1+c) + (z-c)/(1+c)$ on $z$ near $c$, and $e^{\boldsymbol{\alpha X}+\boldsymbol{\beta Y}} \approx e^{\boldsymbol{\alpha}\bar{X}+\boldsymbol{\beta}\bar{Y}}[1 + \boldsymbol{\alpha}(\boldsymbol{X} - \bar{X}) + \beta(\boldsymbol{Y} - \bar{Y})]$ if $Var(\boldsymbol{\alpha}^{\top}\boldsymbol{X}_i)$ and $Var(\boldsymbol{\beta}^{\top}\boldsymbol{X}_i)$ are small.

The previous equation can be approximated by

$$logit(P(X_{j,i} = 1|, y_i, t_i, \mathbf{X}_{-j,i}) \approx y_i\beta_j - y_i e^{\beta_j} + (y_i + 1)\left(\frac{e^{\alpha_j} A(\boldsymbol{X}_{-j,i}) e^{e^{\beta_j} B(X_{-j,i})t^{\gamma}}}{e^{\alpha_j} e^{\sum_{k\neq j}^{r}\bar{X}_k\alpha_k} e^{e^{\beta_j} e^{\sum_{k\neq j}^{r}\bar{X}_k\beta_k} t^{\gamma}} + 1} - \frac{A(\boldsymbol{X}_{-j,i}) e^{B(X_{-j,i})t^{\gamma}}}{e^{\sum_{k\neq j}^{r}\bar{X}_k\alpha_k} e^{e^{\sum_{k\neq j}^{r}\bar{X}_k\beta_k} t^{\gamma}} + 1}\right) + \frac{e^{\alpha_j} A(\boldsymbol{X}_{-j,i})}{1 + e^{\alpha_j} e^{\sum_{k\neq j}^{r}\bar{X}_k\alpha_k}} - \frac{A(\boldsymbol{X}_{-j,i})}{1 + e^{\sum_{k\neq j}^{r}\bar{X}_k\alpha_k}} + \boldsymbol{\sigma}^{\top}\mathbf{X}_{-j} + \sigma_j + \sigma_0 + constant$$

$$\approx y_i\beta_j - y_i e^{\beta_j} + (y_i + 1)\left(\frac{e^{\alpha_j} e^{\sum_{k\neq j}^{r}\bar{X}_k\alpha_k} e^{e^{\beta_j} e^{\sum_{k\neq j}^{r}\bar{X}_k\beta_k} t^{\gamma}}}{e^{\alpha_j} e^{\sum_{k\neq j}^{r}\bar{X}_k\alpha_k} e^{e^{\beta_j} e^{\sum_{k\neq j}^{r}\bar{X}_k\beta_k} t^{\gamma}} + 1}\left[1 + \sum_{k\neq j}^{r}(\bar{X}_k - X_{k,i})\alpha_k + e^{\beta_j} e^{\sum_{k\neq j}^{r}(\bar{X}_k - X_{k,i})\beta_k}\right] - \frac{e^{\sum_{k\neq j}^{r}\bar{X}_k\alpha_k} e^{e^{\sum_{k\neq j}^{r}\bar{X}_k\beta_k} t^{\gamma}}}{e^{\sum_{k\neq j}^{r}\bar{X}_k\alpha_k} e^{e^{\sum_{k\neq j}^{r}\bar{X}_k\beta_k} t^{\gamma}} + 1}\left[1 + \sum_{k\neq j}^{r}(\bar{X}_k - X_{k,i}) + e^{\sum_{k\neq j}^{r}(\bar{X}_k - X_{k,i})\beta_k}\right]\right) + \frac{e^{\alpha_j} e^{\sum_{k\neq j}^{r}\bar{X}_k\alpha_k}}{1 + e^{\alpha_j} e^{\sum_{k\neq j}^{r}\bar{X}_k\alpha_k}}\left[1 + \sum_{k\neq j}^{r}(\bar{X}_k - X_{k,i})\alpha_k\right] - \frac{e^{\sum_{k\neq j}^{r}\bar{X}_k\alpha_k}}{1 + e^{\sum_{k\neq j}^{r}\bar{X}_k\alpha_k}}\left[1 + \sum_{k\neq j}^{r}(\bar{X}_k - X_{k,i})\alpha_k\right] + \boldsymbol{\sigma}^{\top}\mathbf{X}_{-j} + \sigma_j + \sigma_0 + constant$$

This equation is a linear combination of $Y, \exp(\boldsymbol{X}_{-j}) * H_{00}(T), Y * \boldsymbol{X}_{-j}, \boldsymbol{X}_{-j}$.

## 7. Data Availability Statement

Summary data that support the findings of this study are available on request from the senior author. Individual data are not publicly available due to privacy or ethical restrictions.

## References

1. Boag JW. Maximum likelihood estimates of the proportion of patients cured by cancer therapy. *Journal of the Royal Statistical Society*. 1949;11(1):15-53.

2. Peng Y, Yu B. Chapter 2: The parametric cure models. Cure Models: Methods, Applications and Implementation. 2021; 1: 5–38.

3. Yilmaz YE, Lawless JF, Andrulis IL, Bull SB. Insights from mixture cure modeling of molecular markers for prognosis in breast cancer. *Journal of Clinical Oncology.* 2013; 31: 2047–2054.

4. Andrulis IL. *et al.* Neu/erbB-2 amplification identifies a poor-prognosis group of women with node-negative breast cancer. *Journal of Clinical Oncology*. 1998; 16: 1340–1349.

5. Mulligan AM, Pinnaduwage D, Bull SB, O'Malley FP, Andrulis IL. Prognostic effect of basal-like breast cancers is time dependent: Evidence from tissue microarray studies on a lymph node-negative cohort. *Clinical Cancer Research.* 2008; 14: 4168–4174.

6. Feeley LP, Mulligan AM, Pinnaduwage D, Bull SB, Andrulis IL. Distinguishing luminal breast cancer subtypes by Ki67, progesterone receptor or TP53 status provides prognostic information. *Modern Pathology.*2014; 27: 554–561.

7. Forse CL, *et al.* Elevated expression of podocalyxin is associated with lymphatic invasion, basal-like phenotype, and clinical outcome in axillary lymph node-negative breast cancer. *Breast Cancer Research and Treatment.* 2013; 137: 709–719.

8. Meng X. Multiple-imputation inferences with uncongenial sources of input. *Biometrika.* 1992; 9: 538–558.

9. White IR, Royston P, Wood AM. Multiple imputation using chained equations: Issues and guidance for practice. *Statistics in Medicine.* 2010; 30: 377–399.

10. Bartlett JW, Seaman SR, White IR, Carpenter JR. Multiple imputation of covariates by fully conditional specification: Accommodating the substantive model. *Statistical Methods in Medical Research.* 2015; 24: 462–487.

11. Bonneville EF, Resche-Rigon M, Schetelig J, Putter H, Wreede LC. Multiple imputation for cause-specific Cox models: Assessing methods for estimation and prediction. *Statistical Methods in Medical Research.* 2022; 31: 1860–1880.

12. Beesley LJ, Bartlett JW, Wolf GT, Taylor JMG. Multiple imputation of missing covariates for the Cox proportional hazards cure model. *Statistics in Medicine.* 2016; 35: 4701–4717.

13. Clements L, Kimber AC, Biedermann S. Multiple imputation of composite covariates in survival studies. *Statistical Methods in Medical Research.* 2022; 5: 358–370.

14. Van Buuren S. 6.4 Derived variables. *Flexible imputation of missing data* 2nd Ed. Chapman & Hall/CRC Interdisciplinary Statistical Series; 2012

15. Xu C. Improving mixture cure modelling of multiple molecular factors in cancer prognosis. PhD thesis. *University of Toronto Tspace*; 2023 https://tspace.library.utoronto.ca/handle/1807/138482

16. Xu C, Bull SB. Penalized maximum likelihood inference under the mixture cure model in sparse data. *Statistics in Medicine.* 2023; 42: 2134–2161.

17. Heinze G, Ploner M, Beyea J. Confidence intervals after multiple imputation: Combining profile likelihood information from logistic regressions. *Statistics in Medicine.* 2013; 32: 5062–5076.

18. Rubin, R. B. Inference and missing data (with discussion). *Biometrika.* 1976; 63: 581–592.

19. Raghunathan TE, Lepkowski JM, van Hoewyk J, Solenberger P. A multivariate technique for multiply imputing missing values using a sequence of regression models. *Survey Methodology.* 2001; 27: 85–95.

20. Van Buuren S, Groothuis-Oudshoorn CGM. MICE: Multivariate imputation by chained equations in R. *Journal of Statistical Software.* 2011; 45: 1–67.

21. Beesley LJ, Taylor JMG. A stacked approach for chained equations multiple imputation incorporating the substantive model. *Biometrics.* 2021; 77(4):1342-1354.

22. Robert CP, Casella G. Chapter 9: The two stage Gibbs sampler. *Monte Carlo Statistical Methods* 2; 2004

23. Peng Y, Taylor JMG. Residual-based model diagnosis methods for mixture cure models. *Biometrics.* 2017; 73: 495–505.

## Figures and Tables

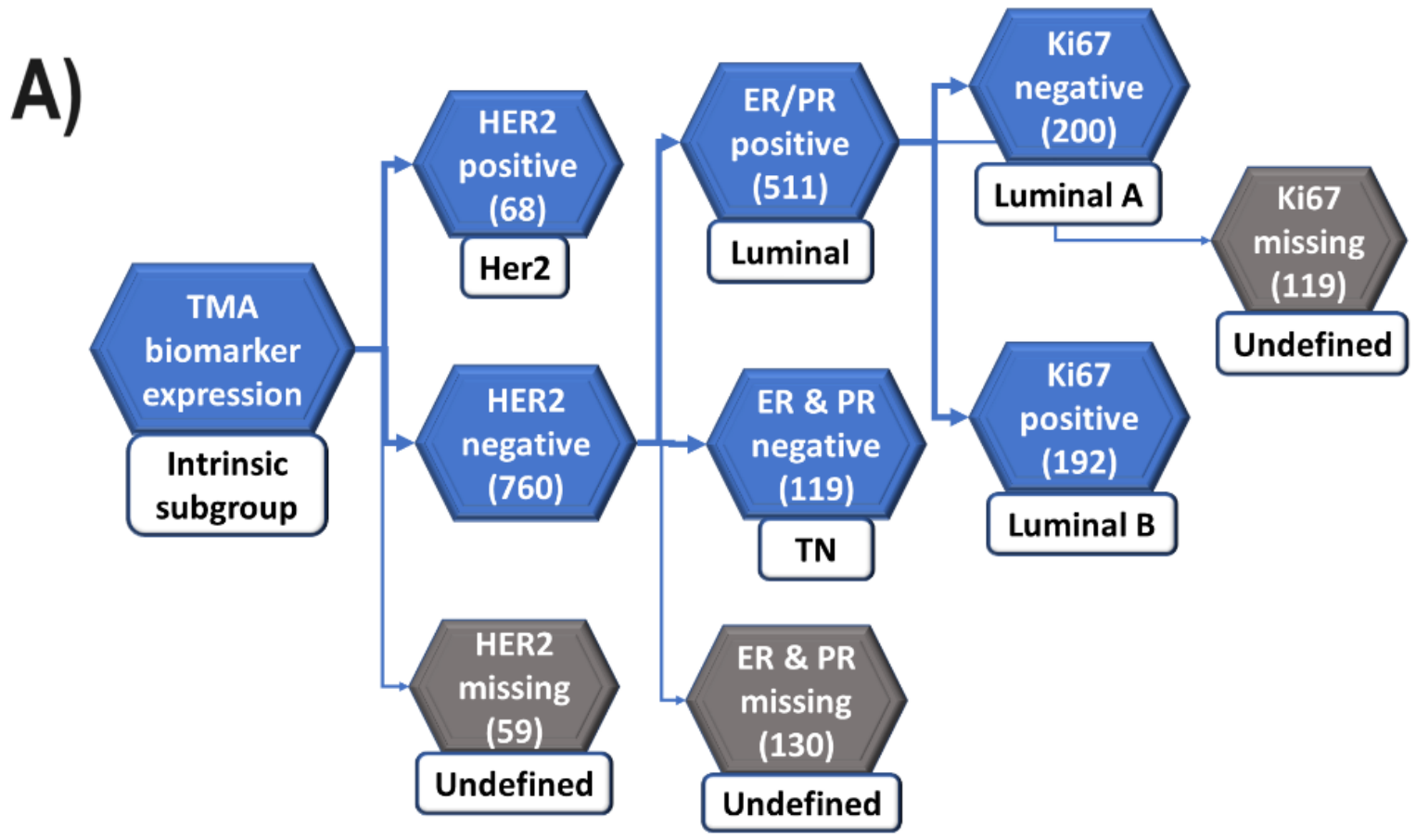

| Biomarker | | Positive | Negative | Missing |
|---|---|---|---|---|
| Her2 | Count | 68 | 760 | 59 |
| | (%) | 7.7% | 85.7% | 6.7% |
| ER | Count | 509 | 205 | 173 |
| | (%) | 57.4% | 23.1% | 19.50% |
| PR | Count | 408 | 321 | 158 |
| | (%) | 46% | 36.2% | 17.8% |
| Ki67 | Count | 400 | 268 | 219 |
| | (%) | 45.1% | 30.0% | 24.9% |

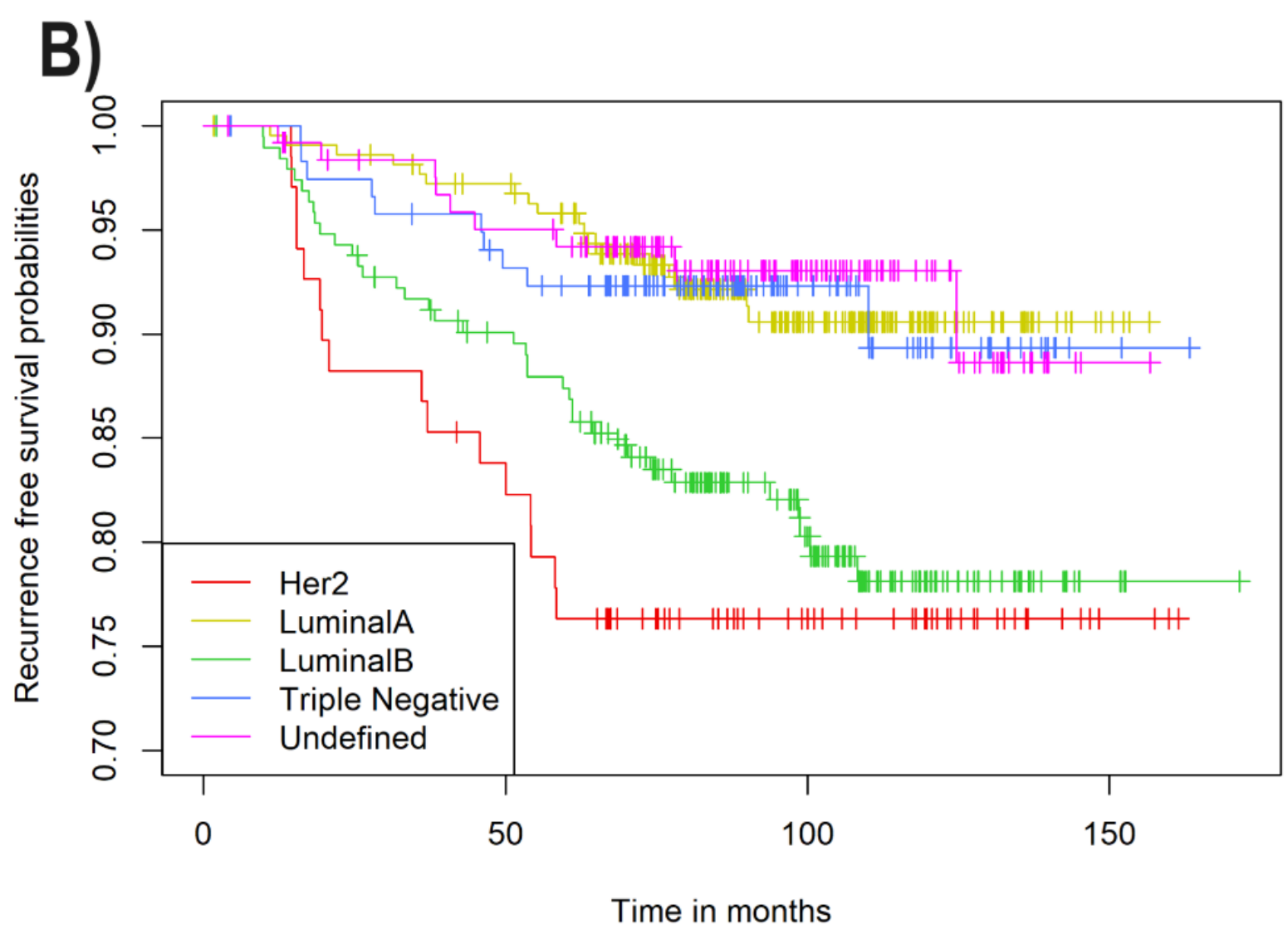

| Subtype | Sample Size | (%) | No. of events |
|---|---|---|---|
| Her2 | 68 | 7.7% | 16 |
| Luminal A | 200 | 22.5% | 18 |
| Luminal B | 192 | 21.6% | 37 |
| TN | 119 | 13.4% | 10 |
| Undefined | 308 | 34.7% | 9 |

Figure 1. A) Schematic to define ANN breast cancer tumour subtypes using TMA biomarker with a side table for counts and percentages of biomarker values and missingness and B) tumour subtype stratified KM plots for TMA data (n=887) with a side table for sample size and number of events for tumour subtypes.

Table 1: Parameter true values in the simulation setup based on MC models with varying covariates. Sample size n=1000, simulation replicates = 2000. Model 1 refers to a 5-variable MC model with 25% or 10% event rate, and Model 2 refers to a 3-variable MC model with 25% event rate.

| Parameter | 25% Event rate | | | | 10% Event rate | |
|---|---|---|---|---|---|---|
| | Model 1 ($\gamma = 1.84$) | | Model 2 ($\gamma = 1.84$) | | Model 1 ($\gamma = 1.84$) | |
| | α | β | α | β | α | β |
| Intercept | -1.1 | -7.56 | 0.8 | -7 | 0.1 | -7.56 |
| Her2 | 0.37 | 0.51 | -0.75 | 1.4 | 0.37 | 0.51 |
| Luminal A | 0.82 | -0.62 | -0.65 | 0.47 | 0.82 | -0.62 |
| TN | 1.1 | 0.5 | -0.2 | 1.4 | 1.1 | 0.5 |
| MENS0 | 0.25 | 0.77 | | | 0.25 | 0.77 |
| TUMCT | -0.61 | -0.03 | | | -0.61 | -0.03 |

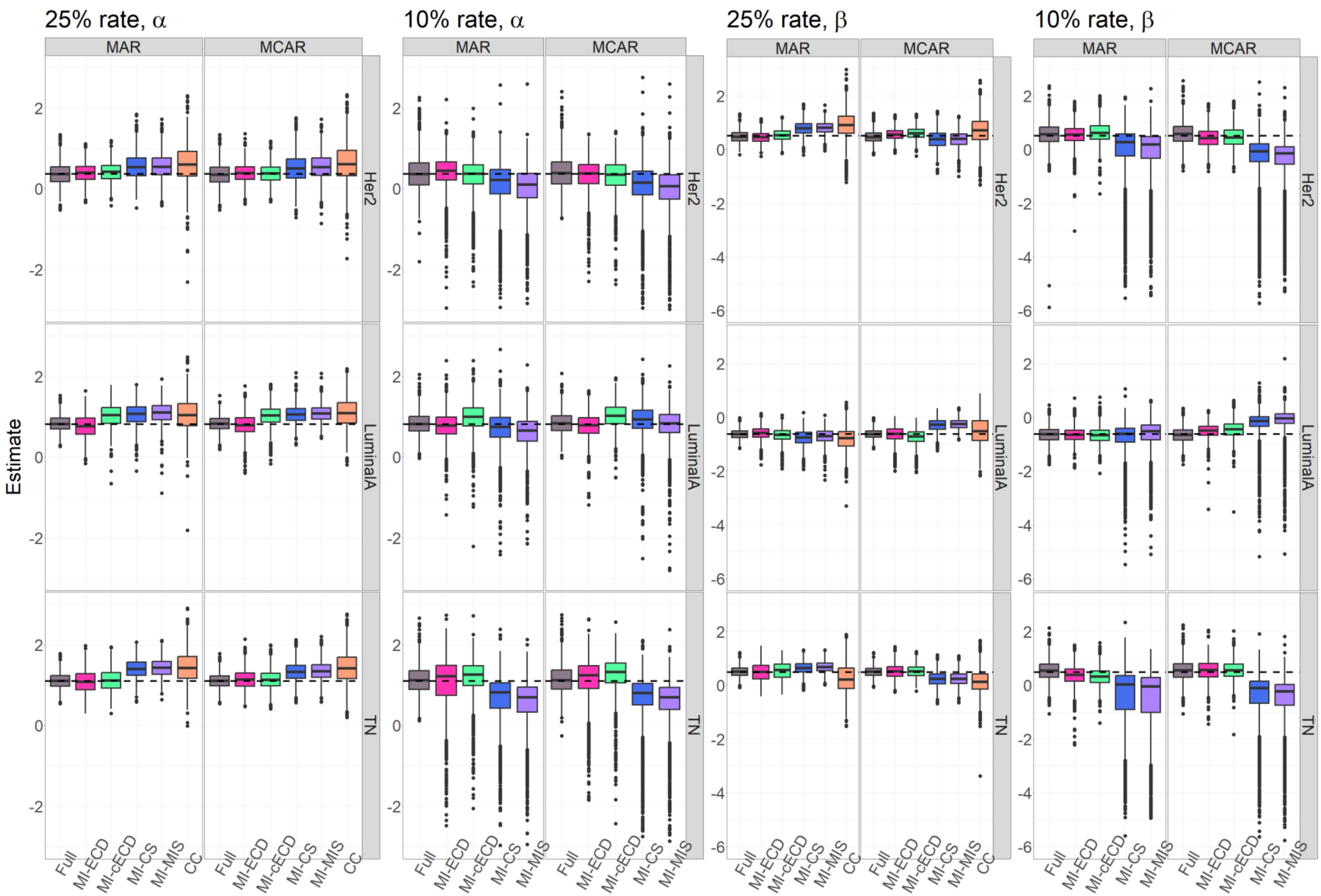

Figure 2. Simulation results under MC Model 1 (2000 replicates). Boxplots of individual parameter estimates for all replicates by FT-PL (sample size n=1000) based on a 5-covariate MC (Model 1) with Her2, LuminalA, TN, MENS0 and TUMCT, for full data (no missingness),

complete case data and MI data with CS, MIS, cECD and ECD imputation models under MCAR or MAR with 25% or 10% event rate. Note that dashed lines indicate generating values of the parameters.

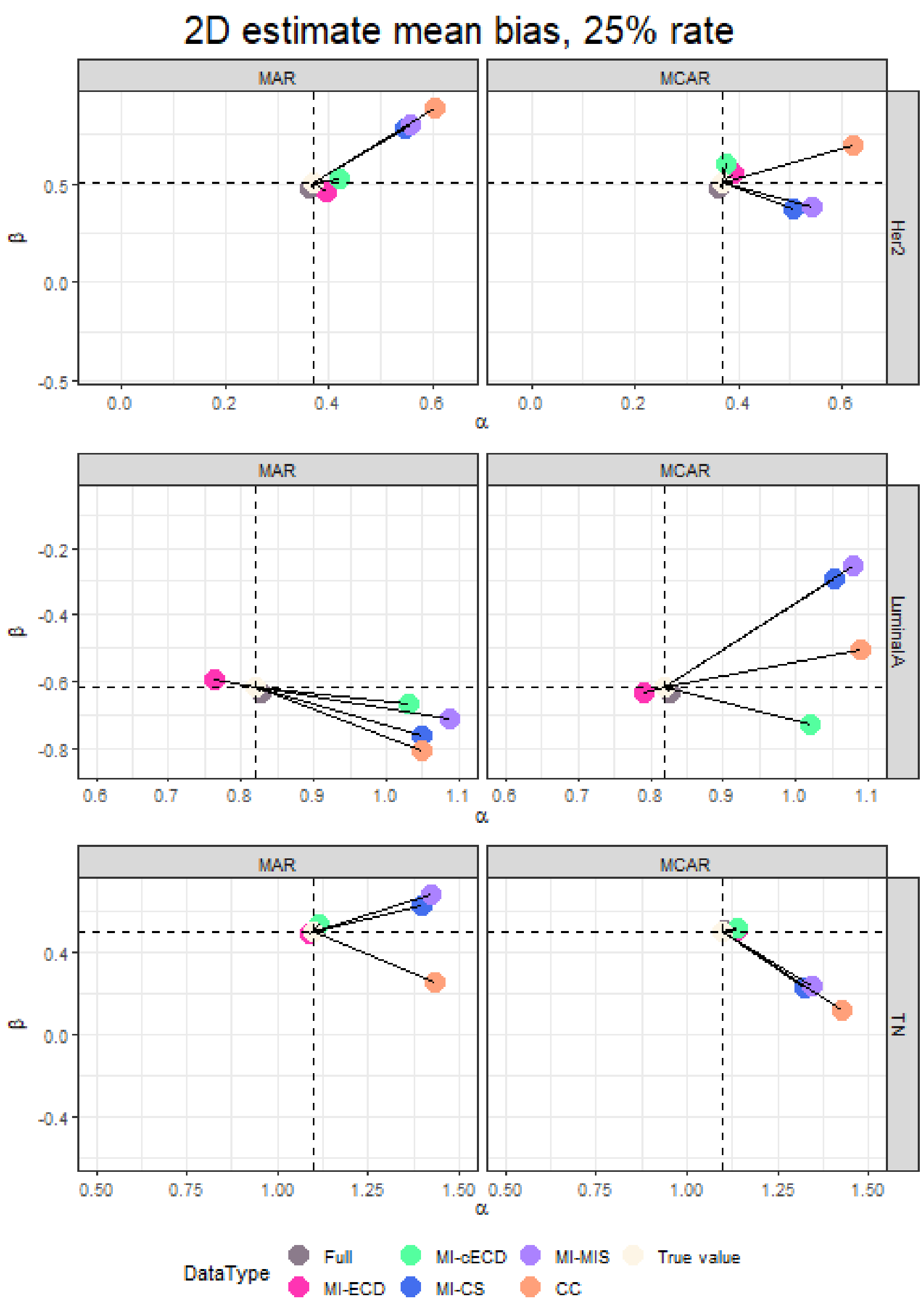

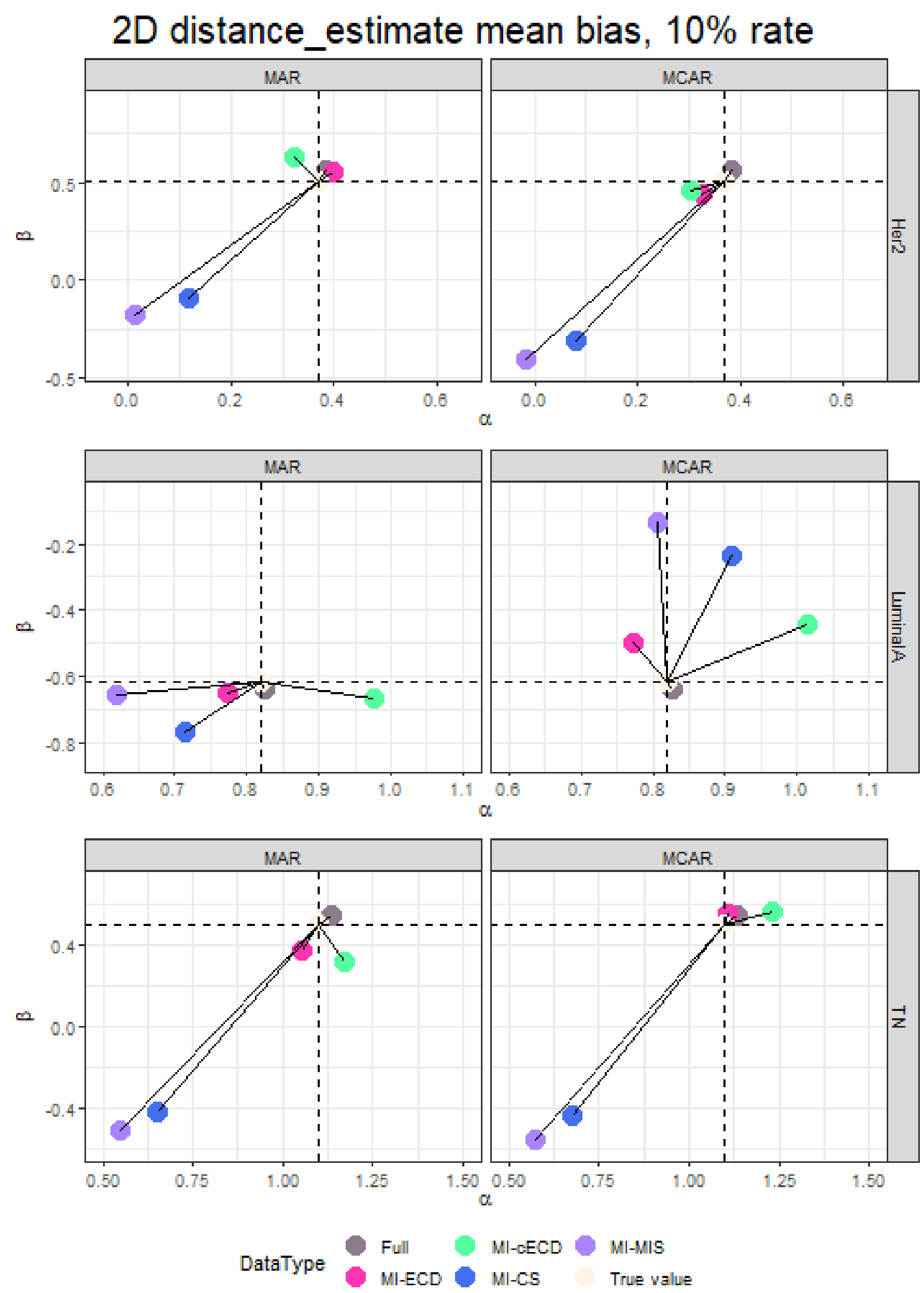

Figure 3. Simulation results under Model 1 (2000 replicates): 2D-distance plots for means of FT-PL estimates of single covariate associated $\alpha$ and $\beta$ estimates, for simulations in Figure 2, under full data (no missingness), complete case data and MI data with several imputation models, under MCAR or MAR with 25% or 10% event rate.

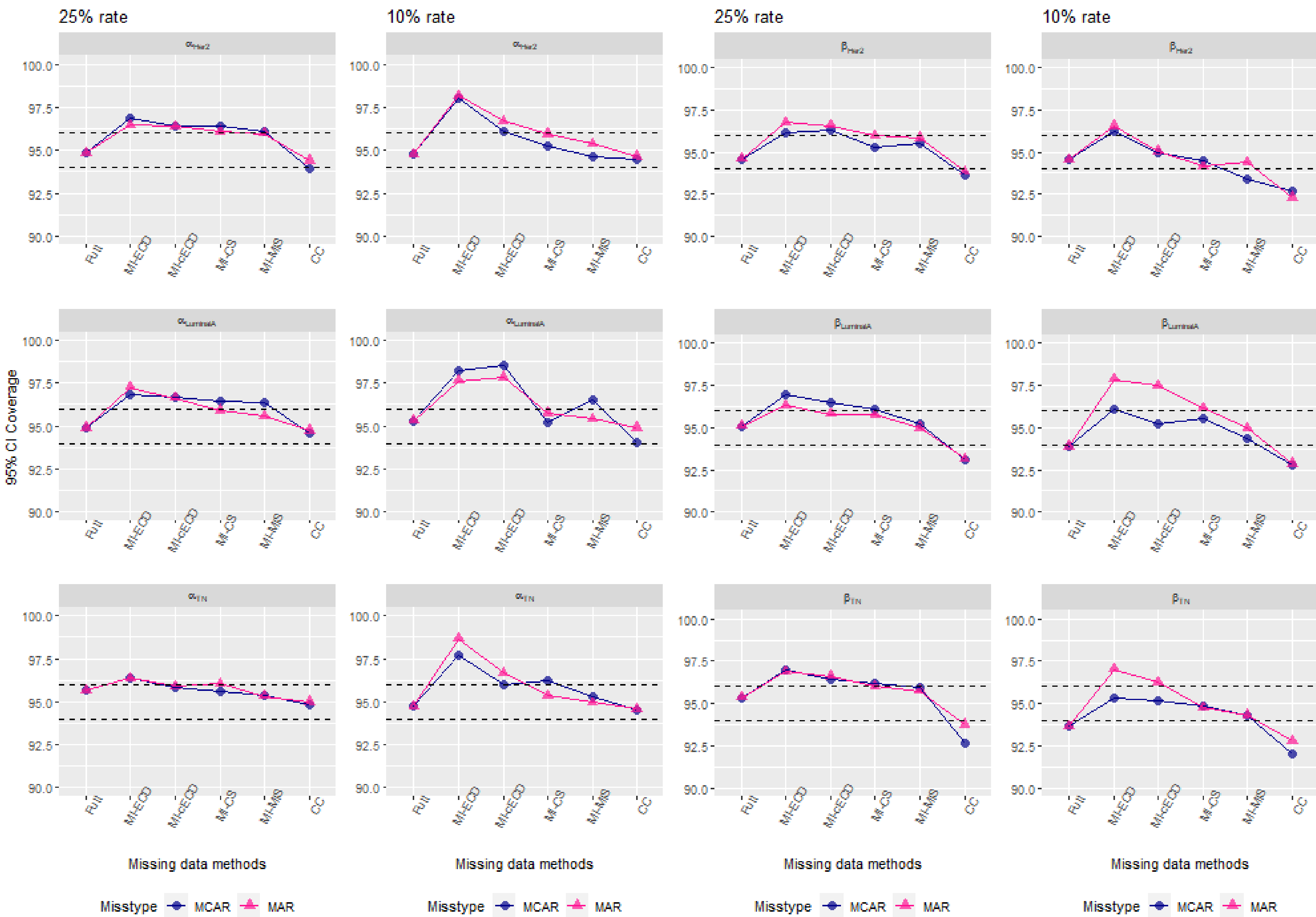

Figure 4. Simulation results under MC Model 1 (2000 replicates): 2-sided PLCI or CLIP-CI coverage rate for estimation by FT-PL, in full data, CC, and MI (with MCAR or MAR), via data based on 5-covariate MC analysis with sample size 1000, under alternative hypotheses at 25% or 10% event rate. Circle denotes CI coverage obtained under MCAR and triangle denotes CI coverage obtained under MAR.

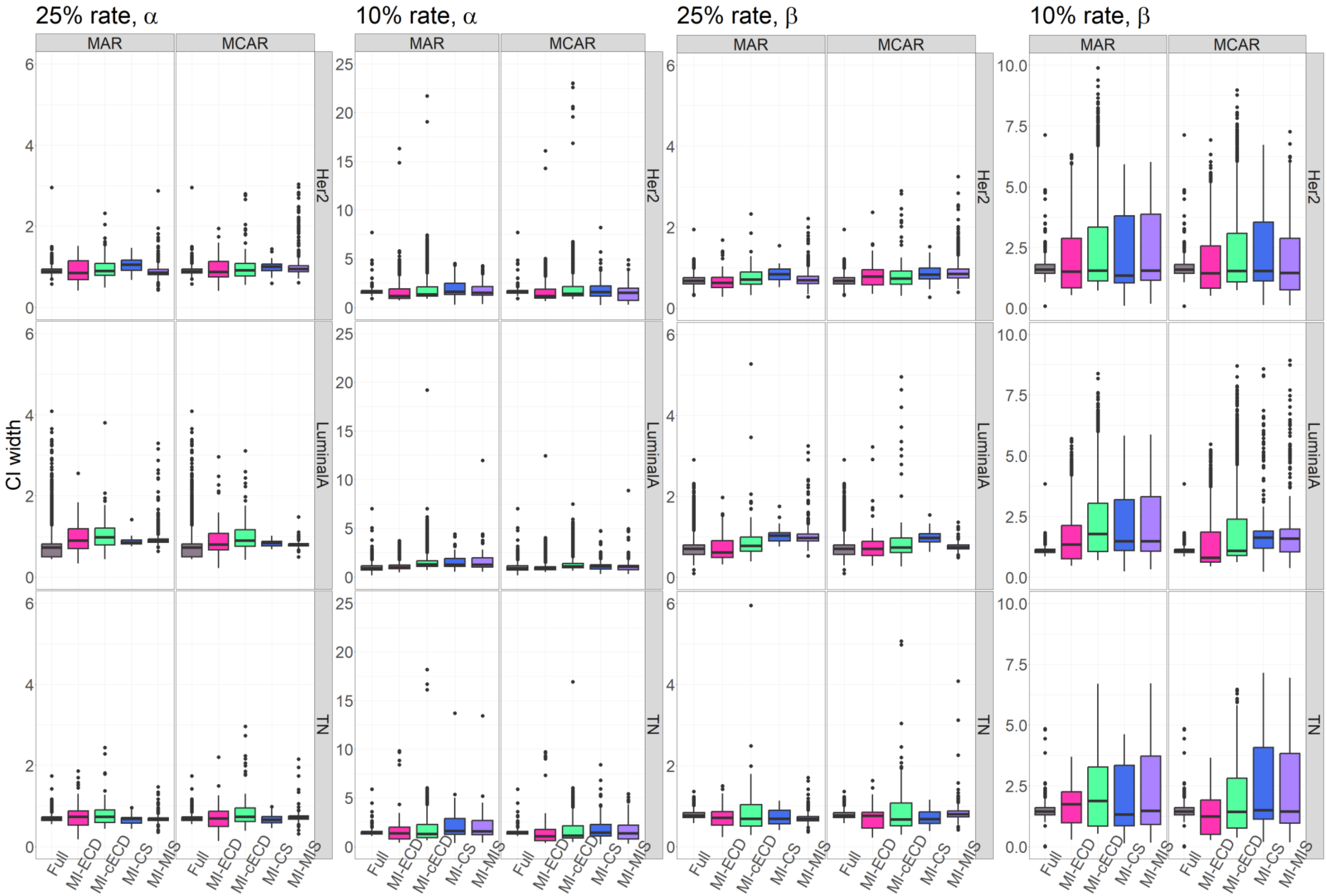

25% rate, α
10% rate, α
25% rate, β
10% rate, β
MAR
MCAR
Her2
LuminalA
TN
CI width
Full
MI-ECD
MI-cECD
MI-CS
MI-MIS

Figure 5. Simulation results under Model 1 (2000 replicates): Boxplots of 2-sided CI widths of PLCI for FT-PL in full data and CI width of CLIP-CI for FT-PL in MI data (with MCAR or MAR), for data based on 5-covariate MC with sample size 1000, under alternative hypotheses at 25% event rate and 10% event rate

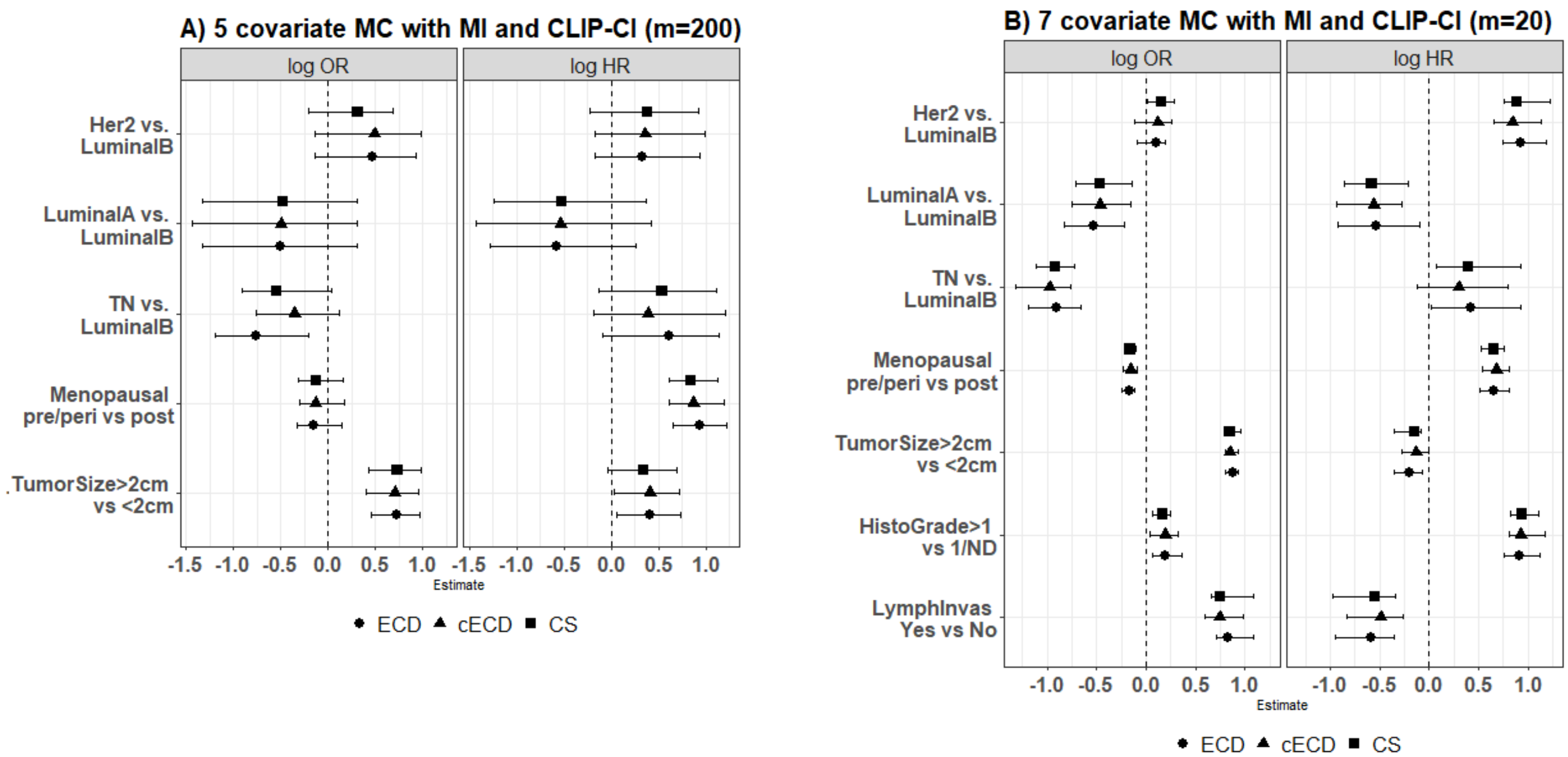

| | p-value (OR) | | | p-value (HR) | | |
|---|---|---|---|---|---|---|
| | ECD | cECD | CS | ECD | cECD | CS |
| Her2 vs. Luminal B | 0.085 | 0.086 | 0.081 | 0.106 | 0.112 | 0.117 |
| Luminal A vs. Luminal B | 0.143 | 0.168 | 0.145 | 0.135 | 0.147 | 0.138 |
| TN vs. Luminal B | 0.009 | 0.096 | 0.052 | 0.114 | 0.124 | 0.117 |
| Menopausal pre/peri vs. post | 0.201 | 0.248 | 0.21 | 0.001 | 0.002 | 0.002 |
| Tumor Size >2cm vs. <2cm | 0.002 | 0.005 | 0.004 | 0.043 | 0.048 | 0.051 |

| | p-value (OR) | | | p-value (HR) | | |
|---|---|---|---|---|---|---|
| | ECD | cECD | CS | ECD | cECD | CS |
| Her2 vs. Luminal B | 0.28 | 0.13 | 0.056 | 4.5E-04 | 8.2E-04 | 3.4E-04 |
| Luminal A vs. LuminalB | 0.004 | 8.4E-03 | 9.5E-03 | 0.015 | 0.009 | 0.01 |
| TN vs. Luminal B | 4.8E-04 | 3.7E-04 | 5.2E-04 | 0.038 | 0.3 | 0.029 |
| Menopausal pre/peri vs. post | 0.013 | 0.02 | 0.02 | 7.2E-03 | 7.9E-03 | 9.2E-03 |
| TumorSize >2cm vs. <2cm | 6.5E-04 | 8.0E-04 | 9.0E-04 | 0.032 | 0.048 | 0.035 |
| HistoGrade >1 vs. 1/ND | 0.027 | 0.044 | 0.04 | 6.8E-04 | 5.0E-04 | 6.0E-04 |
| LymphInvasion Yes vs. No | 6.0E-04 | 7.4E-04 | 3.0E-04 | 1.2E-03 | 3.2E-03 | 0.001 |

Figure 6. Application in ANN Prognostic Analysis. Forest plot of parameter estimates for log(OR) and log(HR) and table of CLIP test p-values for A) 5-covariate MC model with the same MI methods (m=200) and B) 7-covariate MC model with MI methods (m=20), with estimation by FT-PL with CLIP-CI for the ANN data (n=887). Note that MC models towards probability of recurrence.

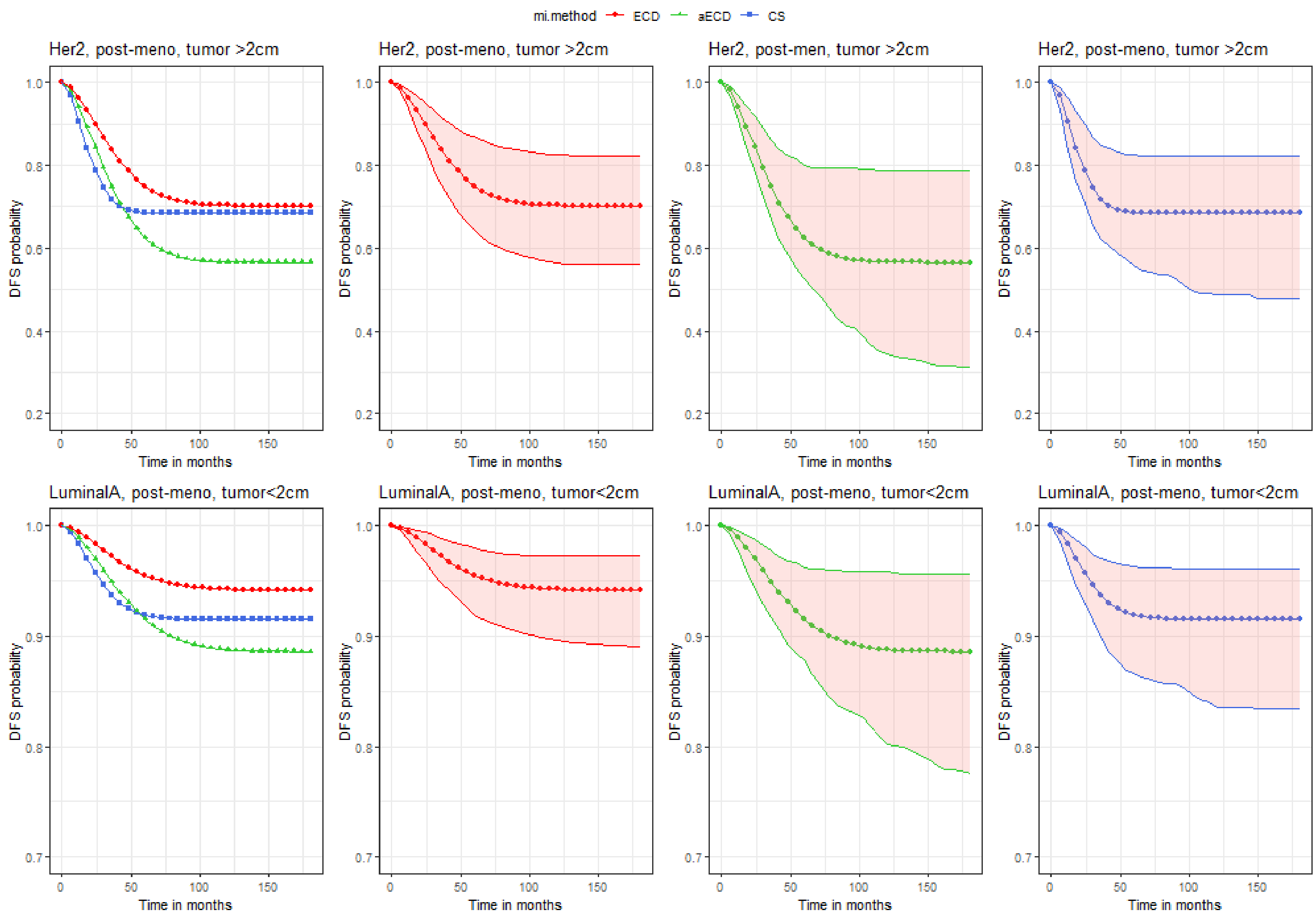

Figure 7. Application in ANN Prognostic Analysis. Predicted DFS probability of hypothetical individuals using estimates by FT-PL under 5-covariate MC (Model 1) with various MI imputation models (m=200). Note that the confidence band is lower 2.5th percentile and upper 97.5th percentile of predicted survival probability among 200 imputations.